# Effect of Substrate on Spin-Wave Propagation Properties in Ferrimagnetic Thulium Iron Garnet Thin Films

Rupak Timalsina,[1] Bharat Giri,[2] Haohan Wang,[2] Adam Erickson,[1] Suchit Sarin,[1] Suvechhya Lamichhane,[2] Sy-Hwang Liou,[2] Jeffery E. Shield,[1] Xiaoshan Xu,[2] and Abdelghani Laraoui[1,2,*]

*[1]Department of Mechanical & Materials Engineering, University of Nebraska-Lincoln, 900 N 16th Street, W342 NH. Lincoln, NE 68588, United States*

*[2]Department of Physics and Astronomy and the Nebraska Center for Materials and Nanoscience, University of Nebraska-Lincoln, 855 N 16th Street, Lincoln, Nebraska 68588, United States*

**Email: alaraoui2@unl.edu*

## Abstract

Rare-earth iron garnets have distinctive spin-wave (SW) properties such as low magnetic damping and long SW coherence length making them ideal candidates for magnonics. Among them, thulium iron garnet (TmIG) is a ferrimagnetic insulator with unique magnetic properties including perpendicular magnetic anisotropy (PMA) and topological hall effect at room temperature when grown down to a few nanometers, extending its application to magnon spintronics. Here, the SW propagation properties of TmIG films (thickness of 7–34 nm) grown on GGG and sGGG substrates are studied at room temperature. Magnetic measurements show in-plane magnetic anisotropy for TmIG films grown on GGG and out-of-plane magnetic anisotropy for films grown on sGGG substrates with PMA. SW electrical transmission spectroscopy measurements on TmIG/GGG films unveil magnetostatic surface spin waves (MSSWs) propagating up to 80 μm with a SW group velocity of 2–8 km $s^{-1}$. Intriguingly, these MSSWs exhibit nonreciprocal propagation, opening new applications in SW functional devices. TmIG films grown on sGGG substrates exhibit forward volume spin waves with a reciprocal propagation behavior up to 32 μm.

## I. Introduction

Spin waves (or magnons) are collective dynamic magnetic excitations found in magnetically ordered materials.[1] They can be generated by injecting current,[2–4] microwave,[5,6] heating up the magnetic material,[7] or with ultrashort laser pulses.[8–10] Some of the key properties of spin waves include long coherence length (up to a few millimeters),[11] high group velocity,[12] high frequencies (up to terahertz),[10,13] and their short wavelength (down to tens of nanometers).[14,15] Other interesting properties of spin waves include nonreciprocal effect,[16] nonlinear effects,[17–19] and multiplexing,[20] making them ideal for low-energy magnetic logical operations and microwave devices.[1] Spin waves in yttrium iron garnet (YIG) have been studied extensively due to the YIG unique magnetic properties including, ultralow magnetic damping ($< 10^{-4}$),[13] long spin-wave coherence length,[5,6,11] and nonlinear magnetic properties.[17–19] Moreover, there is a growing preference for magnetic materials with perpendicular magnetic anisotropy (PMA) due to their ability to efficiently excite electrically spin waves of the material with low current densities,[21–23] relevant to magnon spintronics.[13]

Thulium iron garnet ($Tm_3Fe_5O_{12}$, TmIG), a ferrimagnetic insulator with PMA,[24] has attracted an increasing interest recently for spintronics applications. It can be grown down to a few nanometers[25] while maintaining a low damping.[26] TmIG possesses Dzyaloshinskii-Moriya interaction (DMI) at room temperature[27,28] that can be increased when interfaced with heavy metals (*e.g.*, Pt), leading to topological Hall effect at and above room temperature, which is an indication of the presence of topological spin textures such as skyrmions.[29,30] Studies of spin-wave (SW) propagation properties of TmIG films grown on different substrates are of particular interest for applications in magnon spintronics.[13] Recently, SW electrical measurements of magnetostatic forward volume spin waves (FVSWs) were reported on pristine[31] and bismuth-doped[32] TmIG films with SW velocities and decay lengths of 0.15 – 4.9 km.s$^{-1}$ and ~ 3.2 – 20 μm, respectively. Nitrogen-vacancy (NV) based magnetometry measurements of magnetostatic surface spin waves (MSSWs) were performed on 34 nm thick TmIG grown on GGG, showing a long SW decay length ~ 50 μm and SW wavelength in the range of 0.8 – 2 μm.[33] However, no correlative measurements were reported on the effect of substrate on the SW propagation properties of thin TmIG films.

Here, the SW transport properties of TmIG thin films (thickness of 7 – 34 nm) grown on GGG and sGGG substrates are studied at room temperature. Magnetic measurements by magneto-optical Kerr effect (MOKE), vibrating-sample magnetometer (VSM), and ferromagnetic resonance (FMR) spectroscopy confirm in-plane (IP) magnetic anisotropy of the TmIG films grown on GGG and out-of-plane (OOP) magnetic anisotropy for TmIG films grown on sGGG. In-plane SW electrical transmission measurements on TmIG (7 – 34 nm)/GGG films show MSSWs with a SW group velocity of 2 – 8 km/s and SW decay length of 50 μm. The MSSWs were observed to have nonreciprocal behavior, extending the application of TmIG to microwave devices (*e.g.*, circulators).[34] Out-of-plane SW electrical transmission measurements on TmIG (32 nm)/sGGG films reveal mainly FVSWs with reciprocal behavior, propagating up to 32 μm.

## II. Experimental Results and Discussion

### II.1 Sample growth and structural characterization

TmIG has a ferrimagnetic behavior, originating from the super-exchange interactions between $Tm^{+3}$ (Dodecahedral, Dod), $Fe^{+3}$ (Tetrahedral, Tet), and $Fe^{+3}$ (Octahedral, Oct) sublattices (see Figure 1a). A series of TmIG thin films, with a thickness of 7 – 34 nm (deduced from x-ray reflectometry) were grown on (111) GGG and sGGG substrates, using pulsed laser deposition.[33] Figure 1b shows X-Ray diffraction (XRD) spectra for TmIG films of thicknesses 32 nm and 12 nm grown on GGG and sGGG substrates. For all TmIG film grown on GGG, the TmIG (444) peak is submerged within the GGG (444) peak due to the small lattice mismatch and/or the weak XRD signal obtained on thin (< 15 nm) films. The lattice mismatch between TmIG and sGGG is higher, resulting in distinct XRD peaks of TmIG and sGGG. The atomic force microscopy (AFM) topography images show a smooth surface with atomic terraces and a root mean square (RMS) roughness value of about 0.3 nm for TmIG grown on both GGG (Figure 1c) and sGGG (Figure 1d) respectively over a 4 μm by 4 μm area.

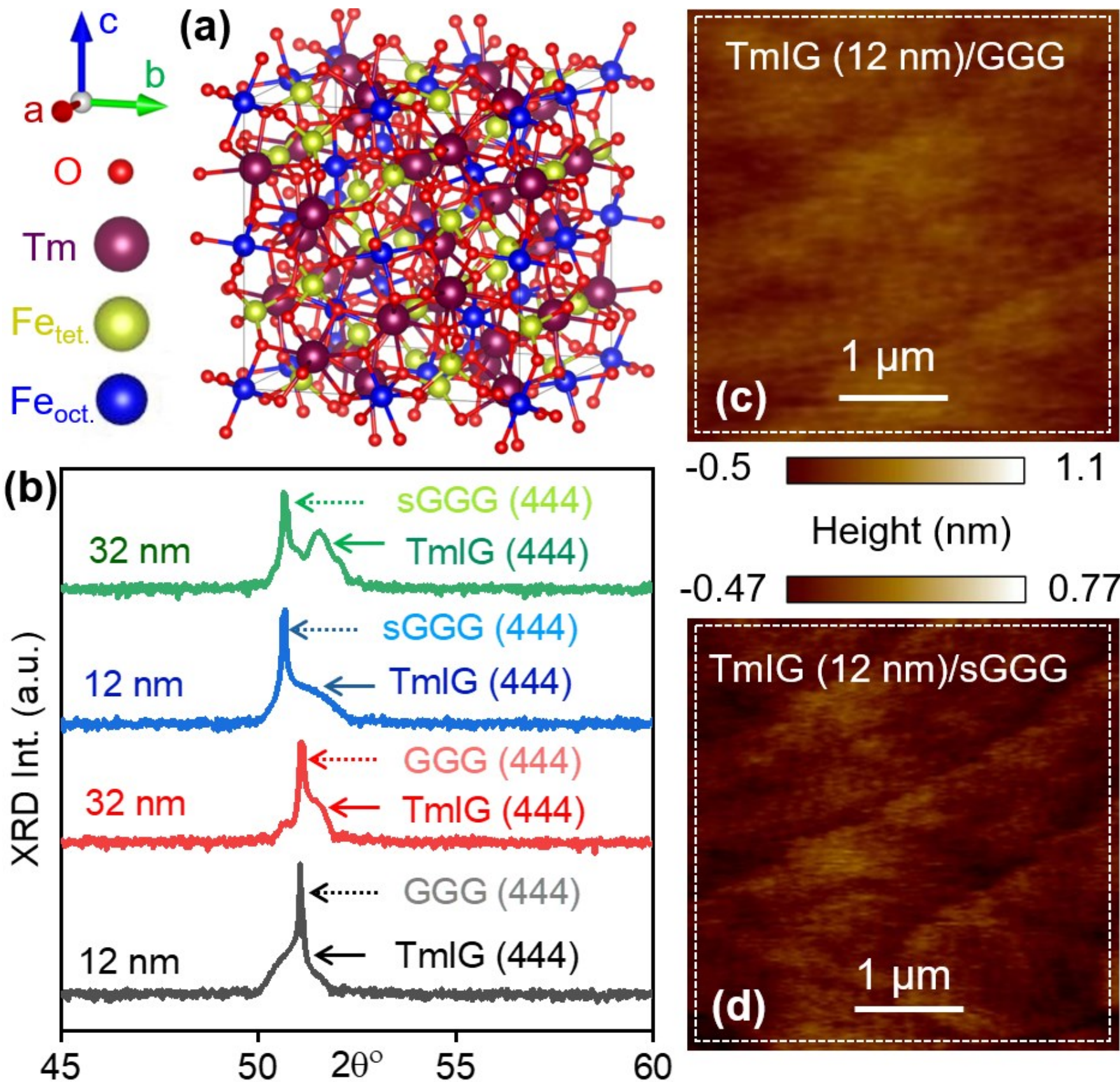

**Figure 1.** Characterization of TmIG films. (a) A schematic of the complex unit cell of TmIG showing tetrahedral Fe (light green) and octahedral Fe (blue) lattice along Tm (purple) and O (red) atoms. (b) Measured XRD spectrum of TmIG films (thickness of 12 – 32 nm) grown on 0.5 mm thick GGG and sGGG substrates. AFM topography image of (c) TmIG (12 nm)/GGG and (d) TmIG (12 nm)/sGGG films.

The TmIG films were also characterized using high-resolution transmission electron microscopy (HR-TEM). The HR-TEM image of TmIG (32 nm)/GGG film (Figure 2a) shows the lattice structure and reveals several dislocations with associated strain consistent with the broad XRD peaks shown in Figure 1b. Scanning TEM (STEM) was used to determine the compositional aspects of the TmIG films. The high-angle annular dark field (HAADF) image in Figure 2b shows sharp contrast between the GGG and TmIG, indicating a lack of intermixing. The layers in the lower left portion of the image are the Pt cover layers to protect the material of interest from Ga ion damage during sample preparation, accomplished via standard lift-out procedures.[35] Energy dispersive x-ray spectroscopy (EDS), in STEM mode, was used to map the elemental distributions of TmIG (32 nm)/GGG film (Figures 2c-h). The EDS maps show a clear delineation between the GGG substrate and TmIG film, with no evidence of interfacial mixing, most evident in the Fe, Tm and Gd maps. Any Ga present in the TmIG was likely due to Ga contamination during ion milling.[35]

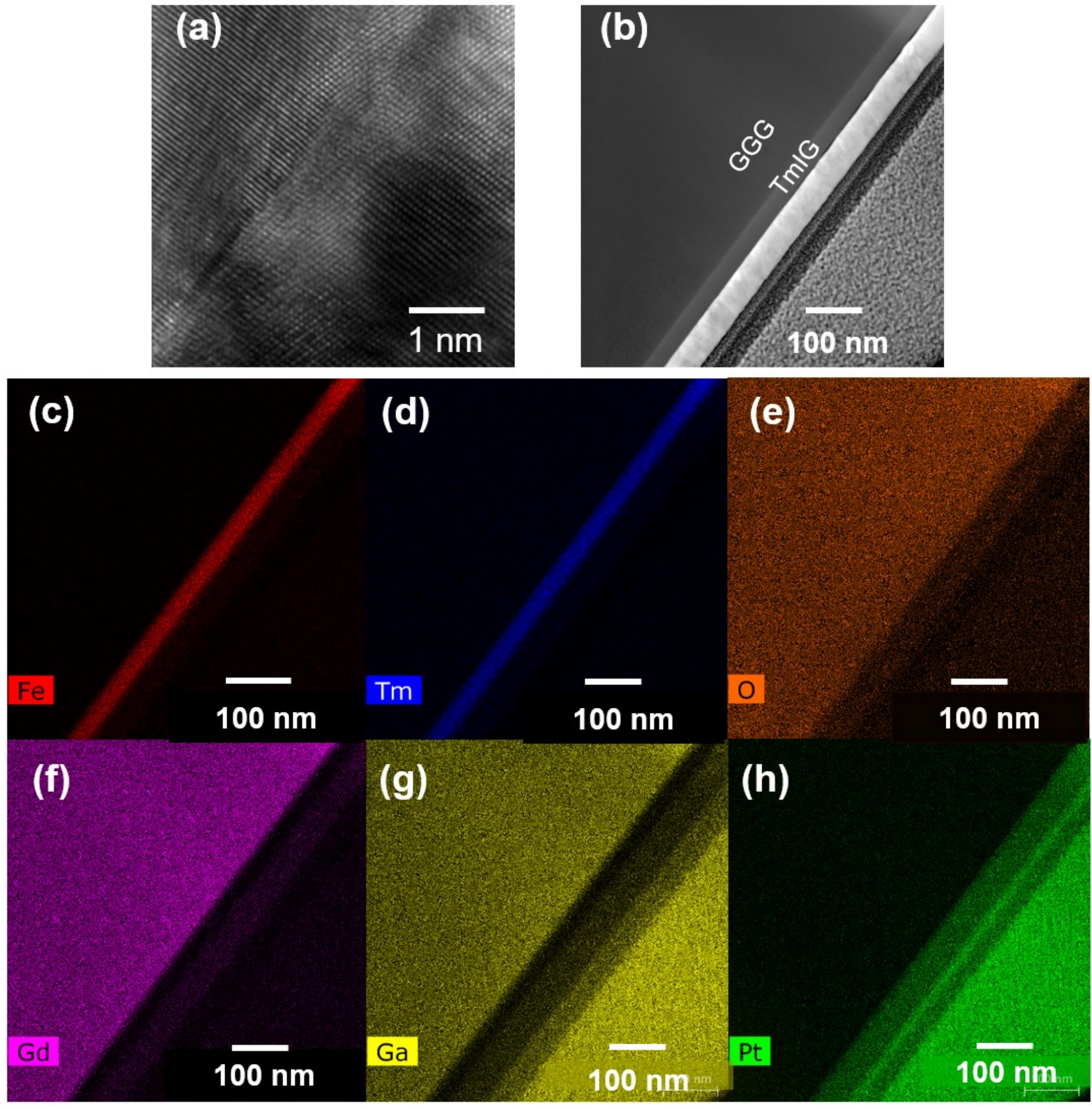

**Figure 2.** (a) HR-TEM image of TmIG (32 nm)/GGG substrate, showing periodic lattice, indicative of excellent crystallinity. (b) Cross-sectional HAADF-STEM image of the same film. (c-h) EDS element maps of Fe, Tm, O, Gd, Ga, Pt respectively, separated by different colors.

Elemental mapping in Figure 2b indicates no interdiffusion between different layers, similar to TmIG films grown by *off-axis* magnetron sputtering.[26] EDS microanalysis was employed to determine the elemental concentrations within the sample. However, due to inherent limitations in EDS, particularly in accurately measuring low atomic number elements, it cannot be used to quantify the oxygen content. To overcome this limitation and obtain accurate elemental ratios, we used X-ray photoelectron spectroscopy (XPS) because of its higher sensitivity and energy resolution, making it well-suited for analyzing the chemical composition of materials, especially for elements like oxygen.[36,37] By employing XPS, the elemental ratios were determined, including the oxygen concentration, with greater precision. See the Supporting information (SI) Section S1 for details of XPS measurements on TmIG (32 nm)/GGG and TmIG (32 nm)/sGGG films. The XPS analysis[37] in SI Figure S1 provides valuable information about the electronic structure and oxidation states of Tm and Fe within the examined material. Linewidth of the Tm, Fe, and O in TmIG/GGG are present in 1.80, 1.10, and 2.22 eV, respectively, which are very close to the desired TmIG stoichiometry, while Tm, Fe, and O in TmIG/sGGG are present in 2.20, 1.20, and 2.50 eV, respectively. The theoretical ratio between the area of octahedral and tetrahedral Fe peaks is 2:3,[38] which is close to the measured ratios of 0.70 and 0.73 for GGG and sGGG, respectively.

### II.2 Magnetic properties of TmIG films

The magnetic anisotropy for all TmIG films grown on GGG and sGGG substrates was investigated by MOKE spectroscopy. Longitudinal MOKE (LMOKE) measurement (Figure 3a) on TmIG (32 nm)/GGG shows hysteresis loop with a saturation field below 5 mT, while polar MOKE (PMOKE) measurement shows a close to square hysteresis loop (inset of Figure 3a) with a saturation field above 100 mT, indicating approximately in-plane magnetic anisotropy. A nearly opposite behavior is observed on the TmIG films grown on sGGG substrates, *i.e.*, the PMOKE loop shows a saturation magnetic field ~ 50 mT with square loop (see Figure 3b) while the LMOKE loop (inset of Figure 3b) shows a saturation field ~ 500 mT, an indication of PMA. The high (weak) PMA in TmIG films grown on sGGG (GGG) correlate well with high (low) tensile strain provided by sGGG (GGG), inducing a sizable out-of-plane magnetocrystalline anisotropy in TmIG films.[39] The magnetic anisotropy properties of the TmIG films are further investigated from IP and OOP VSM and FMR measurements. A similar behavior was observed for 12 nm thick TmIG films grown on both GGG and sGGG substrates and discussed in SI Section S2. PMOKE measurements on TmIG (7 nm)/GGG are discussed in SI Section S2.

To further investigate the magnetic properties of the TmIG films, VSM measurements were conducted at room temperature. Figure 3c shows in-plane *M-H* hysteresis loops of TmIG (32 nm)/GGG and TmIG (12 nm)/GGG films with a low saturation magnetic field ~1 mT. OOP *M-H* hysteresis loop is plotted in the inset of Figure 3c for TmIG (32 nm)/GGG film with a saturation field ~100 mT and saturation magnetization $M_S$ of 66 kA/m. No OOP VSM signal was detected from the TmIG (12 nm)/GGG film due to insufficient sensitivity. The VSM measurements confirm an in-plane easy axis for the TmIG films grown on GGG substrates, consistent with MOKE measurements (discussed above) and previous measurements.[39] Figure 3d displays OOP *M-H* hysteresis loops of TmIG (32 nm)/sGGG and TmIG (12 nm)/sGGG films with a saturation magnetic field ~30 mT and 10 mT, respectively, and $M_S$ of 70 kA/m. Inset of Figure 3d depicts IP *M-H* loops for TmIG (32 nm)/sGGG and TmIG (12 nm)/sGGG with a saturation magnetic field of 1 T and 0.3 T, respectively. The VSM measurements on TmIG/sGGG films confirm out-of-plane easy axis and PMA, consistent with MOKE measurements (discussed above). Temperature dependent magnetization measurements (see SI Figure S3) on TmIG (32 nm)/GGG and TmIG (32 nm)/sGGG films confirm a Curie temperature > 400 K, mainly limited by the temperature range of our VSM magnetometer, see SI Section S2.

FMR spectroscopy is then employed to determine the magnetic properties (magnetic anisotropy and damping) of the TmIG films grown on both GGG and sGGG substrates. A vector network analyzer (Keysight model P5004A) is used to measure the absorption ($S_{11}$ and $S_{22}$ parameters) of the microwave (MW) signal sent through nonmagnetic picoprobes (40A-GSG-150-DP) to coplanar waveguides (CPW) in a ground-signal-ground (GSG) configuration (see Figures 4a and 4b). The CPW were fabricated using direct laser writing (DWL) mask-less optical lithography and subsequent electron-beam deposition of Ti (5 nm)/Au (120 nm). All details of nanofabrication can be found in reference [33].

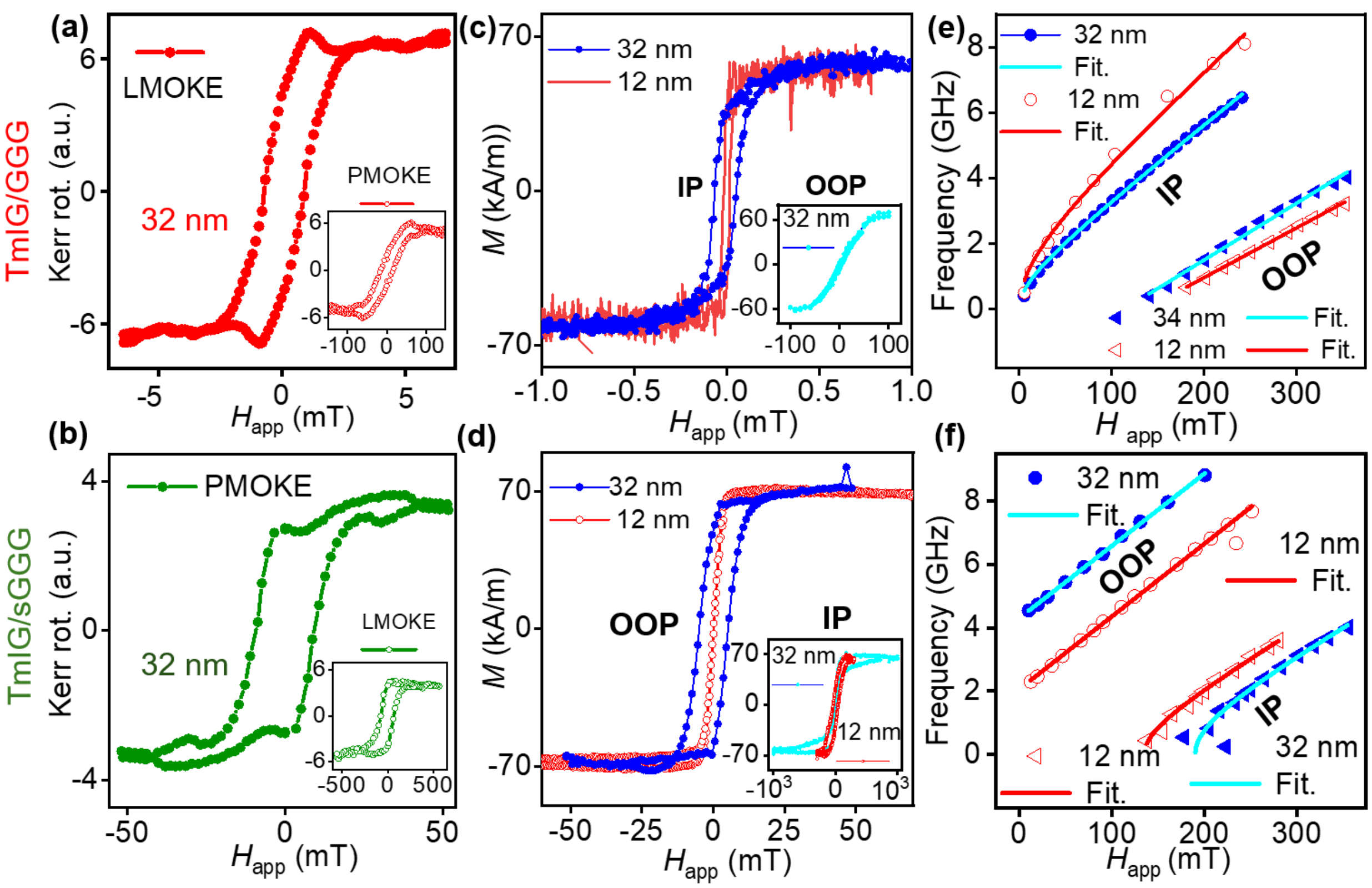

**Figure 3.** Magnetic characterization of TmIG films. (a) LMOKE curve (filled circles) *vs* applied magnetic field $H_{app}$ of TmIG (32 nm)/GGG. Inset of (a) shows PMOKE (open circles) curve *vs* $H_{app}$ of TmIG (32 nm)/GGG. (b) PMOKE (filled circles) curve *vs* $H_{app}$ of TmIG (32 nm)/sGGG. Inset of (b) LMOKE curve (open circles) *vs* $H_{app}$ TmIG (32 nm)/sGGG. (c) IP *M-H* loops of TmIG (32 nm)/GGG (blue filled circles) and TmIG (12 nm) /GGG (red solid line). Inset of (c) OOP loop of TmIG (32 nm)/GGG (sky blue filled circles). (d) OOP *M-H* loops of TmIG (32 nm)/sGGG (blue filled circles) and TmIG (12 nm) /sGGG (red open circles). Inset of (d) IP loop of TmIG (32 nm)/sGGG (sky blue filled circles) and TmIG (12 nm)/sGGG (red open circles). (e) FMR IP curves of TmIG (32 nm)/GGG (blue filled circles) and TmIG (12 nm) /GGG (red open circles), and FMR OOP curves of TmIG (32 nm)/GGG (blue filled triangles) and TmIG (12 nm)/GGG (red open triangles), respectively. (f) FMR OOP curves of TmIG (32 nm)/sGGG (blue filled circles) and TmIG (12 nm) /sGGG (red open circles), and FMR IP curves of TmIG (32 nm)/sGGG (blue filled triangles) and TmIG (12 nm)/sGGG (red open triangles), respectively. All IP and OOP FMR measured curves are fitted with solid lines.[33]

Some of the FMR derivative spectra of the TmIG films are discussed in SI Section S2 and depicted in Figure S4 for 12 nm and 34 nm thick films and Figure S5b for 7 nm thick TmIG/GGG film. Figure 3e shows IP and OOP FMR resonance vs $H_{app}$ for TmIG (32 nm)/GGG and TmIG (12 nm)/GGG films. The resonance magnetic field is higher for OOP than IP measurements on TmIG/GGG films for the same FMR frequency, suggesting in-plane magnetic anisotropy. However, TmIG/sGGG films have the opposite behavior, *i.e.*, the resonance magnetic field is higher for IP than OOP measurements, suggesting PMA.[39,40] The IP and OOP measurements in Figure 3e and Figure 3f were fitted with solid lines[33,39] leading to gyromagnetic ratio $\gamma$ of 22.4 GHz/T and $M_S$ of 66 kA/m for TmIG (32 nm)/GGG film and $\gamma$ of 22.87 GHz/T and $M_S$ of 66 kA/m for TmIG (12 nm)/GGG film. For the TmIG/sGGG films, the linear fit of the OOP measurements in Figure 3f gives $M_s$ = 70 kA/m$^{-1}$ (for both 12 nm and 32 nm films), a negative $H_{app}$-intercept of -198 mT and -99 mT for 32 nm and 12 nm films, respectively. The out-of-plane magnetic anisotropy values are 286 mT and 187 mT for 32 nm and 12 nm thick TmIG/sGGG films. The damping constant $\alpha$ varies in the range of 0.0155 – 0.0365 for IP and OOP FMR measurements done on 32 nm and 12 nm thick TmIG grown on both GGG and sGGG substrates, consistent with previous FMR measurements.[24,26,33,41] The IP FMR measurements performed on TmIG (7 nm)/GGG film (see Figure S5c) revealed $\gamma$ of 22.4 GHz/T, $M_S$ of 66 kA/m (82.9 mT), in plane magnetic anisotropy of 126.8 mT, and $\alpha$ of 0.0122.

### II.3 Study of spin-wave propagation properties in TmIG/GGG films

To measure the spin-wave transport properties of the TmIG films, SW electrical transmission spectroscopy was performed in the MSSW geometry (Figure 4a), *i.e.*, $H_{app}$ is applied in-plane and perpendicular to the SW propagation wavevector. The spin waves are excited by the MW field, injected through the picoprobe to the CPW antennas (Figure 4b) to measure $S_{21}$ and $S_{12}$ parameters. Figure 4c shows the $S_{21}$ intensity map (MW frequency *vs* $H_{app}$) of MSSWs measured at a propagation distance $S$ of 32 μm and at a MW power of 1 mW. One dominating SW mode is clearly seen with a wavevector $k$ of 0.2 rad.μm$^{-1}$, given by the GSG geometry of the stripline.[33] Other SW modes with weak amplitude are visible at high magnetic fields (Figure 4c). The frequency of the main SW mode increases with the applied field $H_{app}$ amplitude as expected from the dispersion curve of the MSSWs.[33] In Figure 4d, selected $S_{21}$ spectra are plotted at $H_{app}$ of 10.65 mT, 20.95 mT, and 40.11 mT, respectively. The frequency difference, $\delta f$, between two neighboring maxima/minima of the $S_{21}$ parameter gives the SW phase of $\pi$, see Figure 4d and Figure S5d in SI section S3 for further details. The SW group velocity $v_g = \frac{\partial \omega}{\partial k} = \delta f . S$,[32] where $S$ is the separation distance between two CPWs (see Figure 4a). Figure 4e depicts the measured SW group velocity versus the applied magnetic field. $v_g$ is in the range of 3 – 7 km/s, slightly higher than the values of 2 – 4 km/s obtained from the TmIG (34 nm)/GGG film (see SI section S3 and Figure S6),[33] explained by the narrower spectral $S_{21}$ peaks for the later one. For TmIG (7 nm)/GGG film, $v_g$ is in the range of 2 to 8 km/s (see Figure S8), like the TmIG (12 nm)/GGG film.

The MSSWs decay length is $l_d = \frac{v_g}{2\pi\alpha f}$,[20] where $f$ is the SW resonance frequency and $\alpha$ is the damping constant obtained from FMR measurements (discussed above). $l_d$ is the range of 20 – 50 μm depending on the amplitude of the applied magnetic field, Figure 4f. This is confirmed from another spectral $S_{21}$ intensity map (MW frequency *vs* $H_{app}$) at a distance $S$ of 80 μm, see Figure 4g. The measured SW amplitudes at $H_{app}$ of 20 mT for four distances $S$ of 32 μm, 40 μm, 60 μm, and 80 μm (Figure S6) are fitted with an exponential decay function (see Figure 4h) with $l_d$ of 30 μm. This value agrees well with the value obtained at $H_{app}$ of 20 mT (see Figure 4f) and are comparable to the values measured on TmIG (32 nm)/GGG films (see SI Section S3). It is difficult to extract $l_d$ from TmIG (7 nm)/GGG film due to the very weak and broad SW intensity, see SI Figure S8. The fabrication of smaller (< 500 nm) CPW antenna to excite efficiency spin waves and usage of scanning NV magnetometry presents a promising avenue for accurately measuring their decay length and wavelengths.[42,43] SW electrical transmission spectroscopy measurement on TmIG (32 nm)/GGG films (Figure S9) show FVSWs with $l_d$ up to 32 μm (SI Section S3), higher than $l_d$ values (≈0.5 μm) obtained in TmIG (60 nm)/NGG substrates.[31,34]

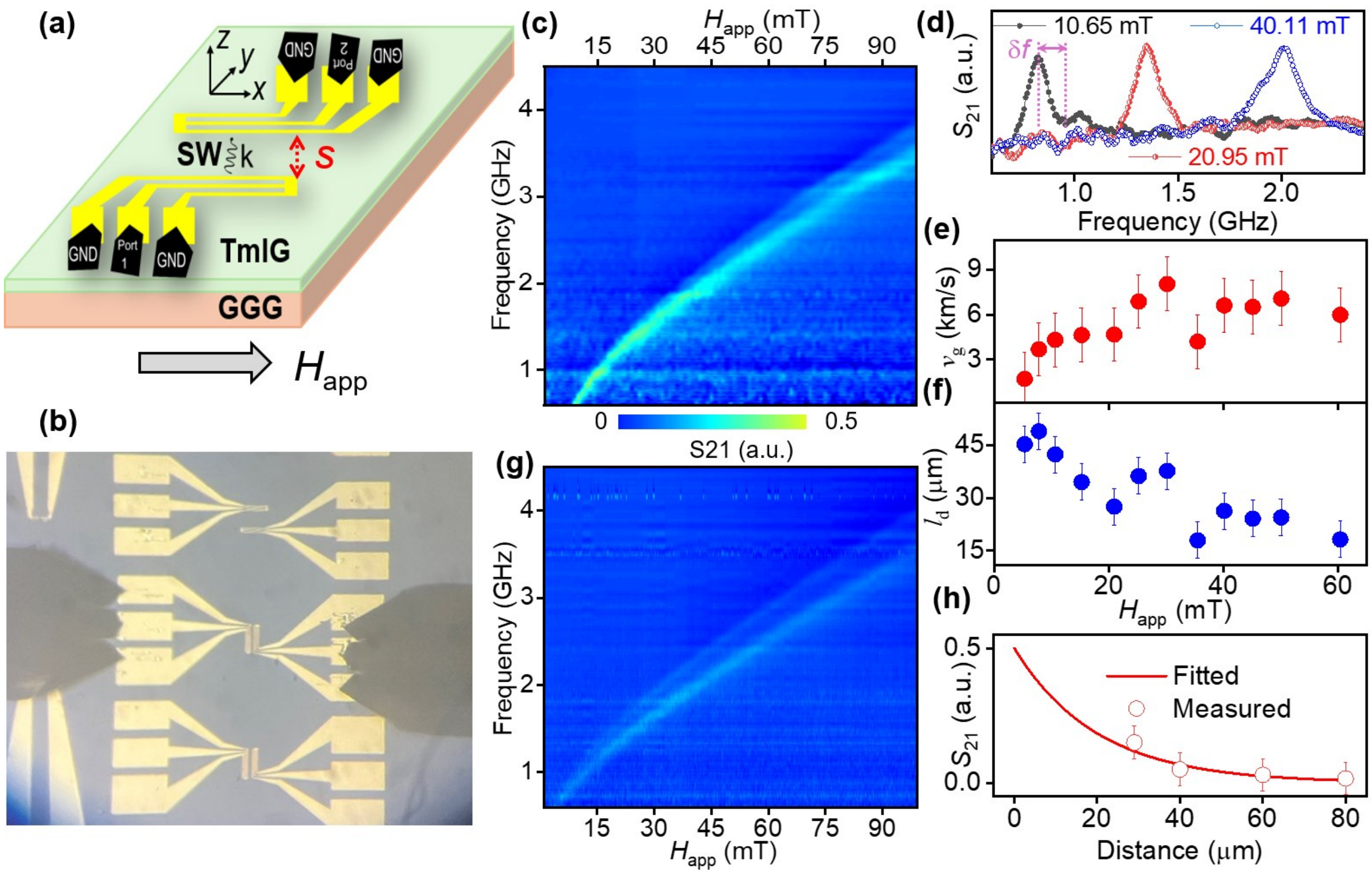

**Figure 4.** Spin-wave electrical transmission spectroscopy in TmIG (12 nm)/GGG films. (a) A schematic of CPWs made on TmIG film. (b) A picture of two nonmagnetic picoprobes connected to the GSG CPW antennas. (c) SW $S_{21}$ intensity map (MW frequency *vs* $H_{app}$) for the MSSW propagation on TmIG (12 nm)/GGG film at a distance $S$ of 32 µm. (d) $S_{21}$ intensity *vs* the MW frequency at $H_{app}$ of 10.65 mT (solid black circles), 20.95 mT (half-filled red circles), and 40.11 mT (open blue circles). SW group velocity (e) and decay length (f) of MSSWs as function of $H_{app}$. (g) $S_{21}$ intensity map (MW frequency *vs* $H_{app}$) of TmIG (12 nm)/GGG at $S$ = 80 µm. (h) Measured (open red circle) and fitted (red solid line) $S_{21}$ amplitude as function of the distance between MW excitation and SW detection. The solid fit in (h) is an exponential decay function with a decay of 30 µm.

### II.4 Spin-wave nonreciprocity properties in TmIG/GGG films

MSSWs in TmIG/GGG films were also found to exhibit nonreciprocal behavior, confirmed by measuring $S_{21}$ and $S_{12}$ parameters. Figures 5a and 5b show $S_{21}$ and $S_{12}$ intensity maps (MW frequency *vs* $H_{app}$) obtained on TmIG (12 nm)/GGG film at a MW power of 1 mW, respectively. Figure 5c displays the line-cut plots of $S_{21}$ and $S_{12}$ spectra obtained from Figures 5a and 5b at $H_{app}$ of 25 mT, applied in plane. No SW signal is observed for $S_{12}$ amplitude *vs* frequency curve, suggesting unidirectional SW propagation. However, $S_{11}$ and $S_{22}$ (reflection) spectra show a similar SW signal (same resonance frequency and amplitude). The nonreciprocity parameter $\beta$ from SW transmission spectra is defined as:[20] $\beta = \frac{S_{21}}{S_{12}+S_{21}}$. Figure 5e shows $\beta$ of 1 as function of $H_{app}$ in the range of 5 – 45 mT, implying a strong nonreciprocal behavior of MSSWs.[16]

The nonreciprocal property is observed when the excited spin waves by the CPW antenna produce stray magnetic fields along $y$ ($H_y$) and $z$ ($H_z$) directions with amplitudes larger in phase compared to the out-of-phase.[44] A similar behavior is observed for TmIG (32 nm)/GGG films (see SI Figure S10). Prior measurements have shown nonreciprocal propagation of spin waves in YIG (100 nm)/GGG[16,45] and permalloy (20 nm) film.[46] This is the first experimental measurement of SW nonreciprocity on TmIG/GGG films, expanding its applications in miniaturized, broad-band, and highly tunable isolators within microwave circuits for spin wave computing, a field where magnons serve as carriers, transporters, and processors of information.[16]

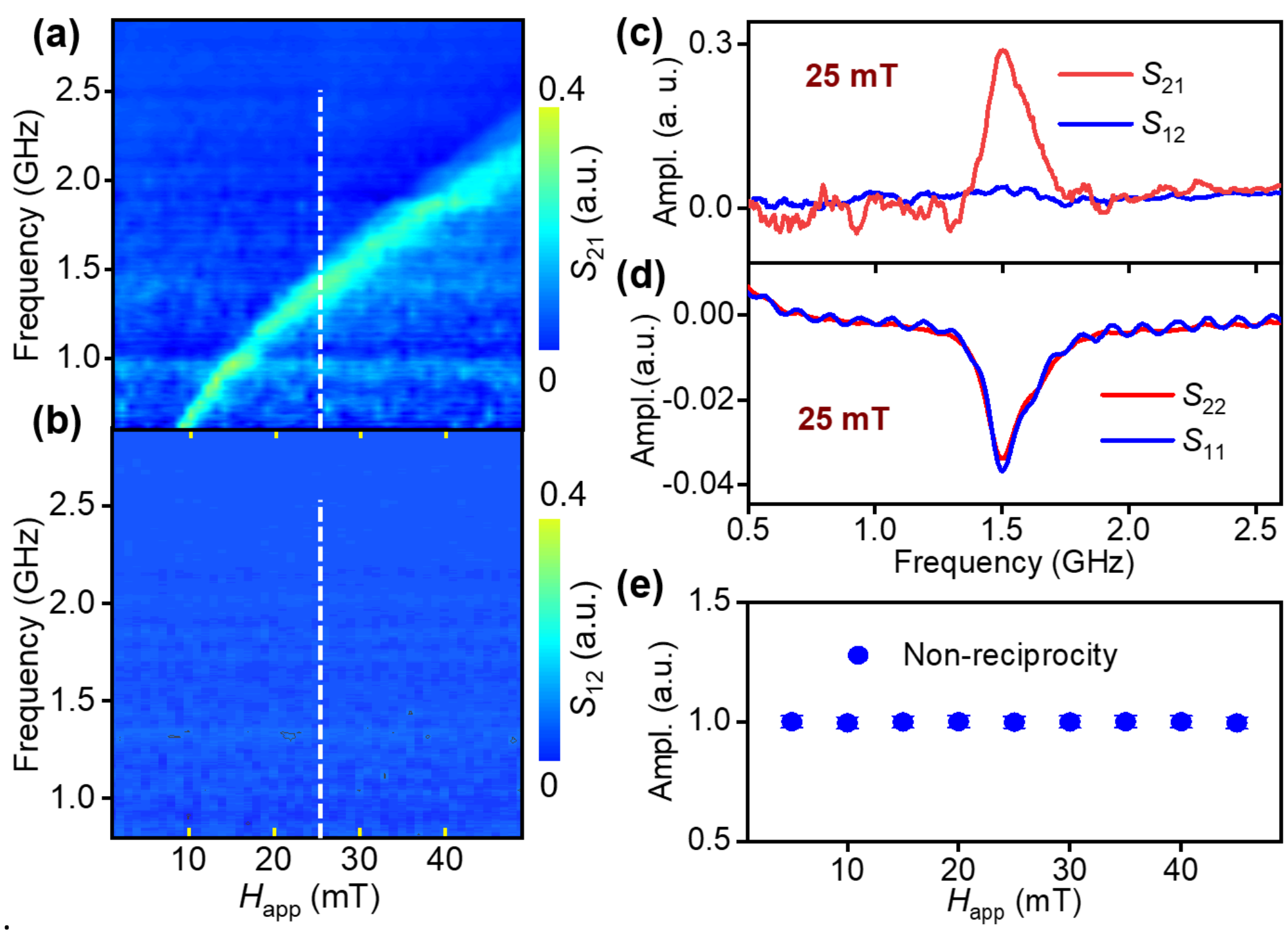

**Figure 5.** Nonreciprocity properties of MSSWs. $S_{21}$ (a) and $S_{12}$ (b) intensity maps (MW frequency *vs* $H_{app}$) on TmIG (12 nm)/GGG films. (c) Line cut plots of $S_{21}$ (red solid line) and $S_{12}$ (blue solid line) at $H_{app}$ of 25 mT. (d) Reflection (FMR) parameters $S_{11}$ (red line) and $S_{22}$ (blue line) on TmIG (12 nm)/GGG measured at $H_{app}$ 25 mT. (e) Nonreciprocity parameter of MSSWs as a function of $H_{app}$.

### II.5 Study of forward volume spin-wave properties in TmIG/sGGG films

TmIG films grown on sGGG substrates have PMA (discussed above), and no SW transmission is observed in plane (*i.e.*, MSSW geometry). Figures 6a and 6b show the SW $S_{21}$ and $S_{12}$ intensity maps (MW frequency *vs* $H_{app}$) of TmIG (32 nm)/sGGG, respectively, measured in FVSW geometry (*i.e.*, magnetic field is applied perpendicular to the sample plane) at a propagation distance $S$ of 32 μm and MW power of 1 mW. The amplitude of the measured FVSW (see Figure 6c at $H_{app}$ of 142 mT) is weak in comparison to the MSSW measured on TmIG (12 – 34 nm)/GGG films (Figure S6). This can be explained by the large separation distance (3.5 μm) between the CPW antenna, compared to the short wavelength (< 500 nm) of the FVSWs, resulting in inefficient MW excitation.[14] Indeed, the $S_{21}$ parameter shows similar resonance and amplitude to $S_{12}$ at 142 mT suggesting a bidirectional propagation of FVSWs. Similar strength and amplitude of $S_{11}$ and $S_{22}$ reflection spectra at 142 mT (see Figure 6d), suggests the proper functioning of our CPW antennas. This marks the first experimental measurements of FVSWs in TmIG/sGGG with PMA. Figure 6e shows almost perfect reciprocal behavior ($\beta$ = 0.5) of the FVSWs over a wide range of applied magnetic fields.

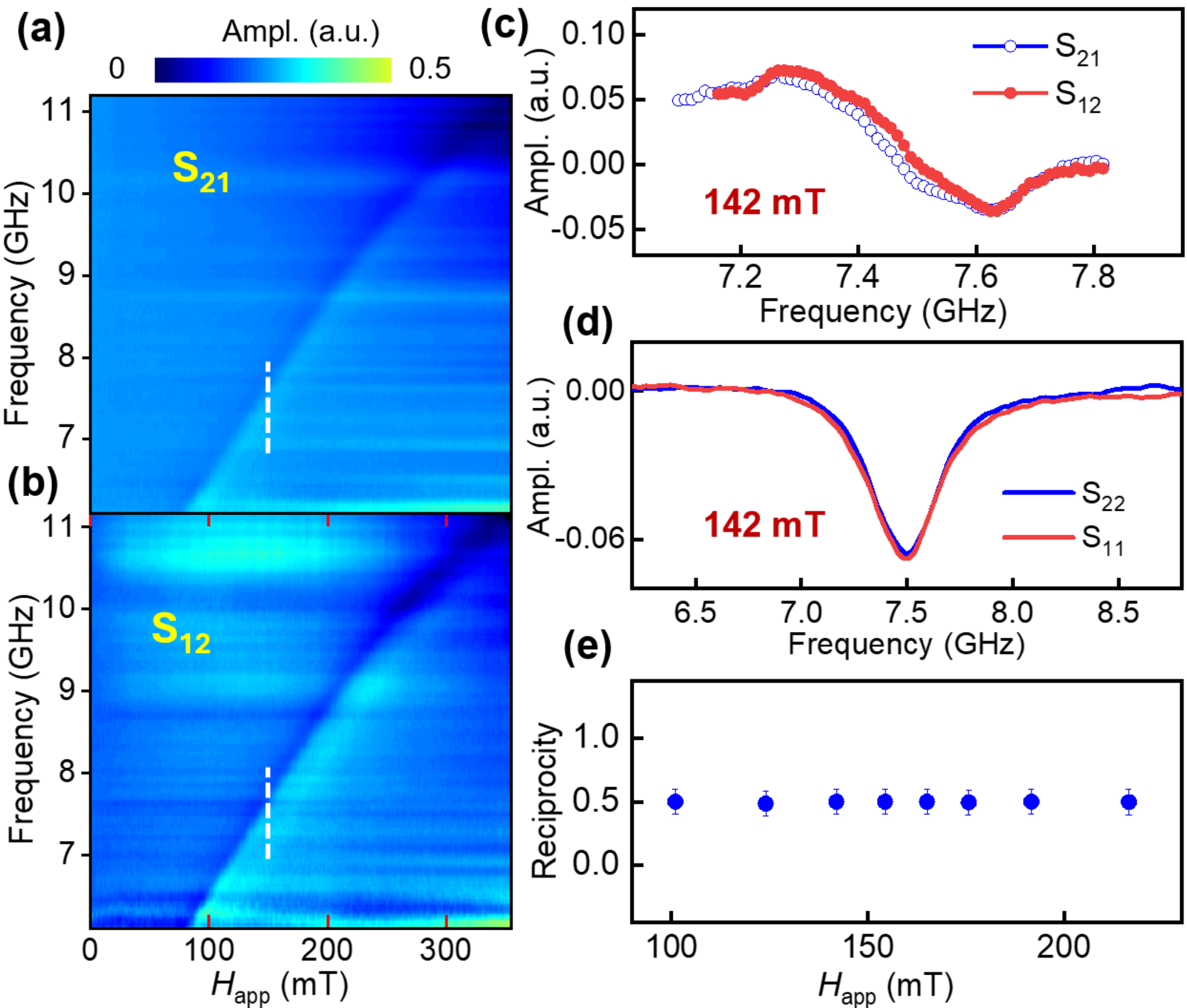

**Figure 6.** $S_{21}$ (a) and $S_{12}$ (b) intensity maps (MW frequency *vs* $H_{app}$) for FVSWs propagation on TmIG (32 nm)/ sGGG at a separation distance $S$ of 32 μm. (c) Line cuts of $S_{21}$ (blue open circles) and $S_{12}$ (open red circles) at 142 mT, extracted from (a) and (b), highlighted by dashed liens. (d) $S_{11}$ (red solid line) and $S_{22}$ (blue solid line) of TmIG (32 nm)/sGGG measured at 142 mT. (e) Reciprocity parameter vs $H_{app}$ of FVSWs.

## III. Conclusions

In summary, spin-wave propagation properties are studied in epitaxially grown TmIG films on GGG and sGGG substrates. Magnetic measurements *via* MOKE, FMR, and VSM confirm in-plane easy axis for TmIG grown on GGG and an out-of-plane easy axis for TmIG grown on sGGG. SW transmission electrical spectroscopy show propagating MSSWs in TmIG/GGG films up to ~ 80 μm and SW velocity of 2 – 8 km/s, depending on the strength of the applied magnetic field. The extended decay lengths signify enhanced SW transport efficiency, offering promising avenues for the development of low loss magnonic devices. Notably, our experimental results reveal nonreciprocal propagation of MSSWs in TmIG/GGG films, suggesting potential applications in nonreciprocal magnonics and logic devices.[45] TmIG films grown on sGGG exhibit isotropic propagation of FVSWs with a long SW decay length of 32 μm.

TmIG has the potential to host topological spin textures,[29] paving the way for investigating new magnetic phenomena such as the interaction between skyrmions and spin waves[47] or topological magnons.[48] The usage of scanning NV magnetometry[43,49,50] may allow mapping of spin waves in TmIG with a short wavelength (< 100 nm), opening new avenues for using spin waves in magnetic insulators in magnon spintronics, [13] and as quantum buses for applications in quantum information processing.[51–53]

**Acknowledgements**

This material is based upon work supported by the NSF/EPSCoR RII Track-1: Emergent Quantum Materials and Technologies (EQUATE) Award OIA-2044049 and NSF award# 2328822. The research was performed in part in the Nebraska Nanoscale Facility: National Nanotechnology Coordinated Infrastructure and the Nebraska Center for Materials and Nanoscience (and/or NERCF), which are supported by NSF under Award ECCS: 2025298, and the Nebraska Research Initiative.

**Author Contributions**

R.T. performed FMR and SW electrical transmission spectroscopy on TmIG films; H.W. grew the TmIG films; B.G. performed MOKE measurements; A.E. assisted R.T. in AFM measurements; S.S. and J.E.S performed HR-TEM, STEM, and EDS measurements; S.L. and S.-Y.L. performed XPS spectroscopy; X.X. and A.L. designed the experiments and supervised the project; R.T. wrote the manuscript with assistance from A.L. and contributions of all authors.

**Conflict of Interest**

The authors declare no conflict of interest.

**Data Availability Statement**

The data that support the findings of this study are available from the corresponding author upon reasonable request.

## Supporting Information

### S.1. X-ray photoelectron spectroscopy characterization on TmIG films

To further check the stoichiometry of the grown TmIG films, high resolution (step size of 0.1 eV) X-ray photoelectron spectroscopy (XPS) was used. Figures S1a-c and Figures S1d-f depict the XPS measured (scattered red lines) spectra of the Tm 4d, Fe 2p, and O 1s peaks of the 32 nm thick TmIG film grown on a (111) GGG and sGGG substrates, respectively. The measured XPS peaks were fitted using a combined Lorentzian and Gaussian function.[37] The atomic percentage analysis was performed using the XPS software.[37] Due to spin–orbit coupling, the Tm 4d and Fe 2p core levels undergo bifurcation into two components in the XPS spectra.[54] The $Tm^{3+}$ $4d_{5/2}$ peak is identified at 176.7 eV, closely resembling a previously documented peak at ~175.9 eV,[38] Additional $Tm^{3+}$ $4d_{3/2}$ peak is observed at 180 eV. The Fe $2p_{3/2}$ peak is situated at 711.2 eV, proximate to the $Fe^{3+}$ peak at 711.0 eV and distant from the $Fe^{2+}$ peak at 709.0 eV. The Fe $2p_{1/2}$ peak is observed at 724 eV. Furthermore, an extra $O_{1s}$ peak at 531.7 eV is attributed to water vapor present on the film's surface.[37]

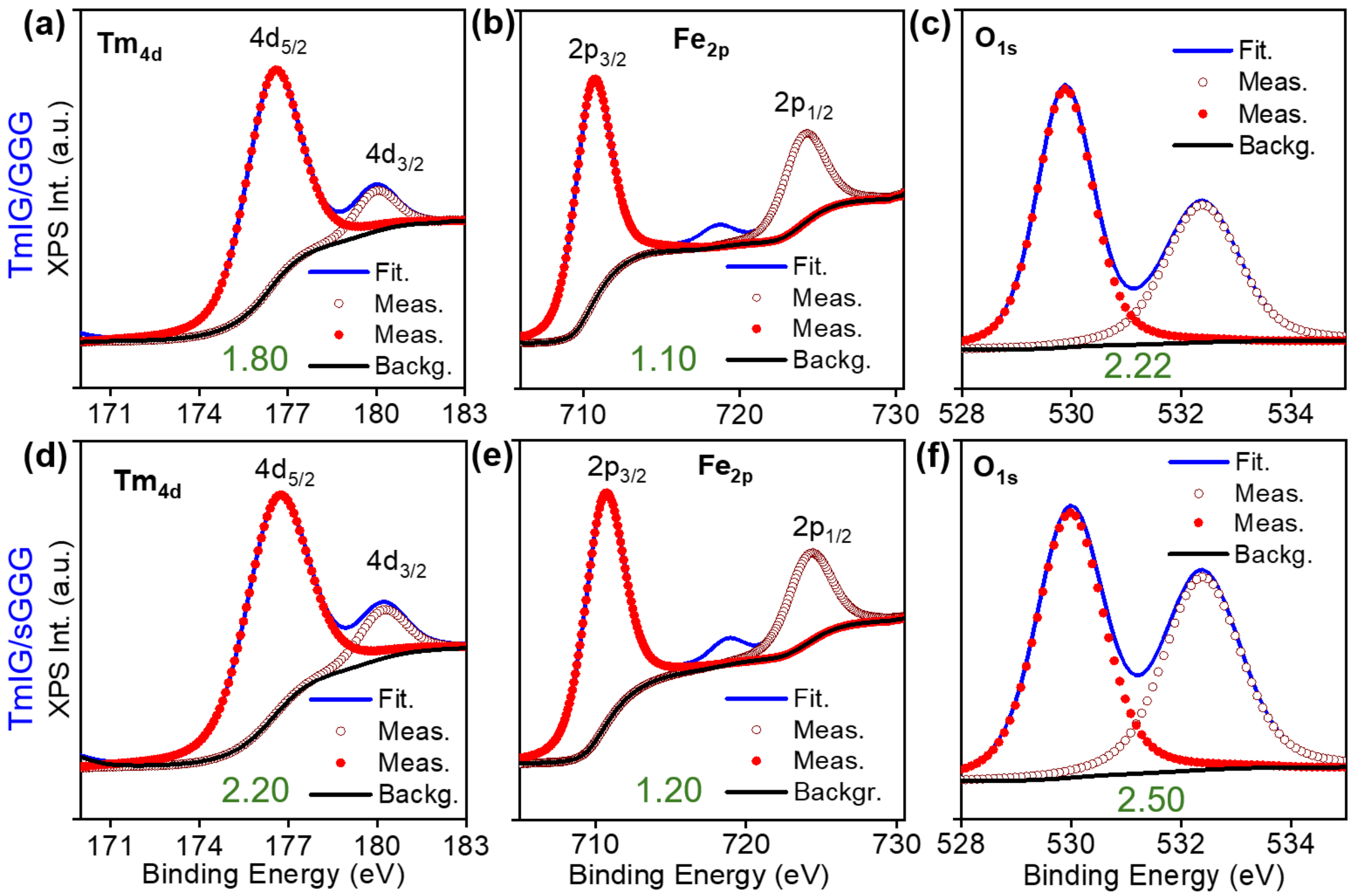

**Figure S1.** High Resolution XPS spectra of TmIG (32 nm)/GGG (a-c) and TmIG (32 nm)/sGGG (d-f). The scattered curves are the measured XPS curves while the solid lines are Gaussian and Lorentzian fits. The black line shows the background coming from inelastic scattered electrons.

### S.2. Additional magnetic characterization on TmIG films

Polar magneto optical (PMOKE) and longitudinal magneto optical (LMOKE) were conducted *vs* the applied magnetic field $H_{app}$ on TmIG (12 nm)/GGG film (Figure S2a). A low saturation magnetic field (~40 mT) was found for LMOKE loops (see the inset of Figure S2a), indicating in-plane magnetic anisotropy as in the case of the 34-nm thick TmIG/GGG films, see the main text.

LMOKE (Figure S2b) on TmIG (12 nm)/sGGG film shows a saturation field ~30 mT, an order of magnitude below the saturation field (~300 mT) for PMOKE loop, indicating out-of-plane magnetic anisotropy and PMA (see the inset of Figure S2b).

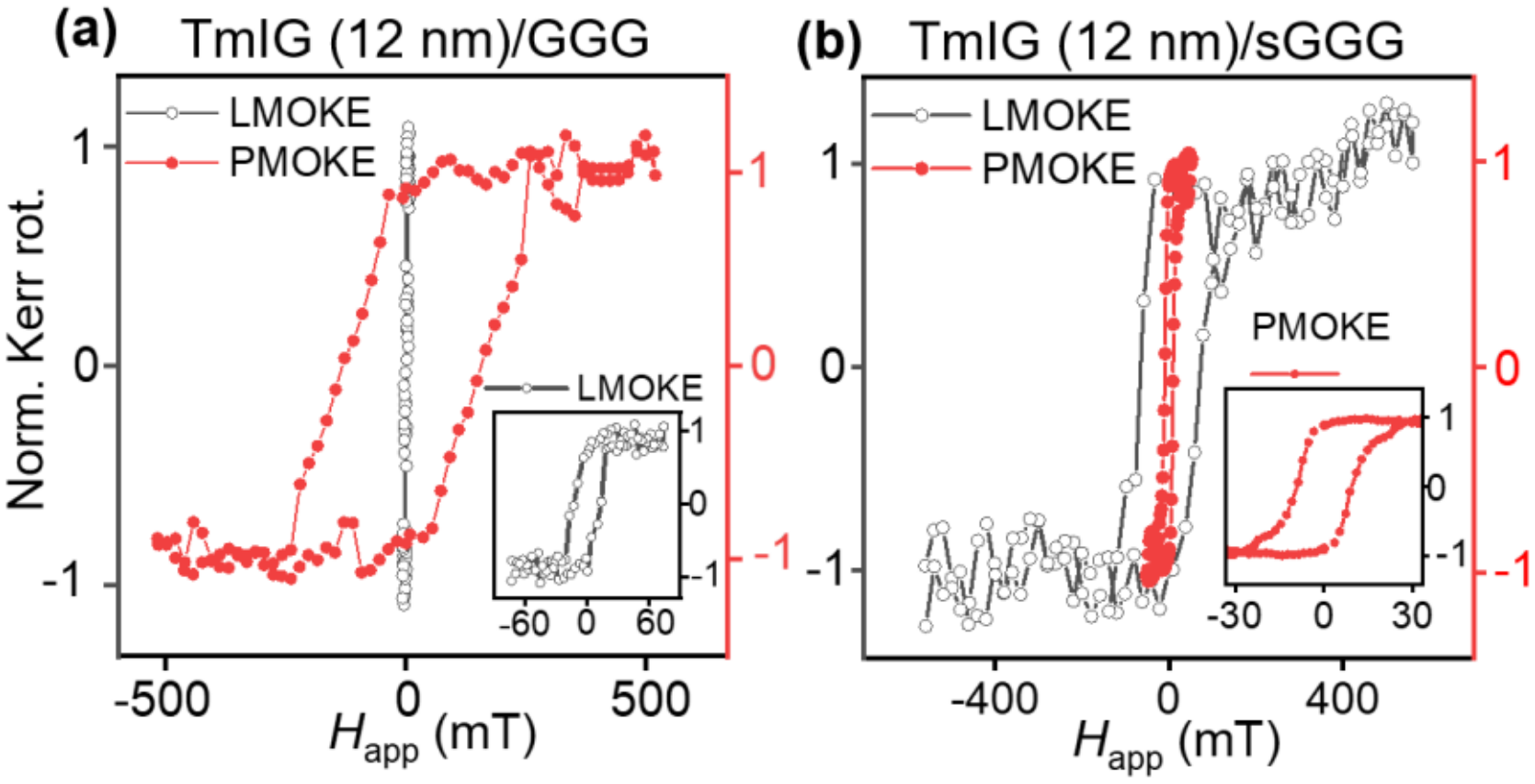

**Figure S2.** (a) Normalized Kerr rotation for LMOKE (open circles) and PMOKE (filled circles) on TmIG (12 nm)/GGG film. Inset of (a): Zoomed LMOKE loop. (b) Normalized Kerr rotation for LMOKE (open circles) and PMOKE (filled circles) of TmIG (12 nm)/sGGG film. Inset of (b): Zoomed PMOKE loop.

To determine the Curie temperature of our TmIG films, the magnetization was measured as a function of temperature using VersaLab 3T cryogen-free vibrating-sample magnetometer (VSM). Figures S3a and S3b show the change in magnetization as a function of temperature for the 32 nm thick TmIG/GGG and TmIG /sGGG films respectively, showing measurable magnetization even at 400 K. Our VSM has the limitation in achieving temperatures > 400 K, suggesting a Curie temperature > 400 K for both TmIG (32 nm) films grown on GGG and sGGG. These measurements are consistent with previous findings.[30]

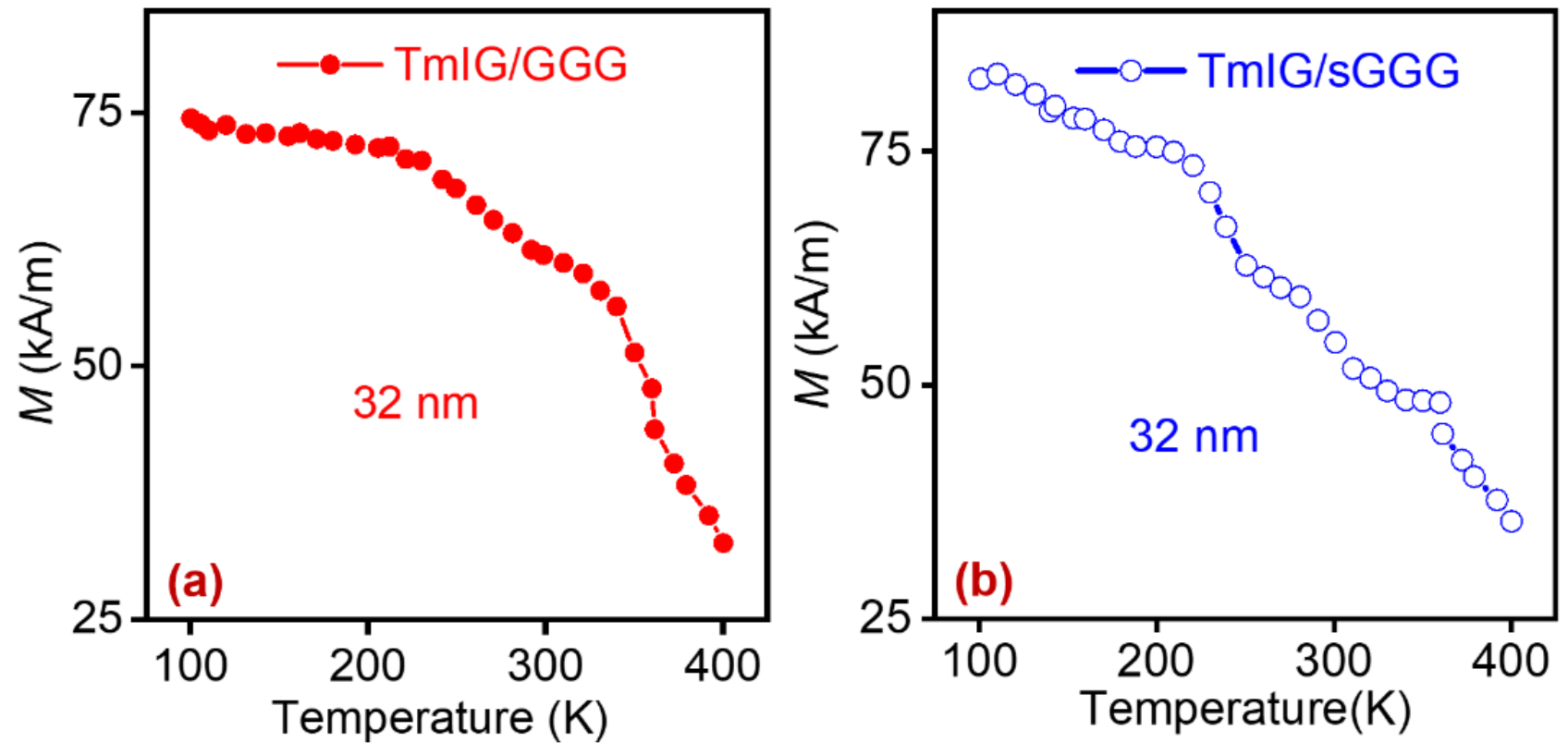

**Figure S3.** Magnetization as function of temperature for TmIG (32 nm)/GGG (a) and TmIG (32 nm)/sGGG (b).

In-plane (IP) and out-of-plane (OOP) ferromagnetic resonance (FMR) were performed by measuring the $S_{11}$ and $S_{22}$ parameters using a vector network analyzer (VNA, Keysight model P5004A) at room temperature, to extract the magnetic properties (magnetic anisotropy and damping constant $\alpha$) of all the epitaxial TmIG films grown on both GGG and sGGG substrates. Figure S4 shows the FMR derivative spectra *vs* $H_{app}$ for various thicknesses (12 nm and 32 nm) of TmIG thin films, showing distinct behaviors at different microwave (MW) frequencies. For IP and OOP measurements on the TmIG grown on GGG, FMR spectra were obtained for MW frequencies ranging from 0.5 to 4 GHz. In contrast, IP and OOP FMR spectra of TmIG grown on sGGG substrates showed MW frequencies in the range of 2 to 5 GHz and 4 to 9 GHz, respectively. Specifically, TmIG films grown on GGG display easy IP axis, while the films grown on sGGG substrates demonstrated easy out of plane magnetocrystalline anisotropy and PMA.[39] Further details can be found in the main text.

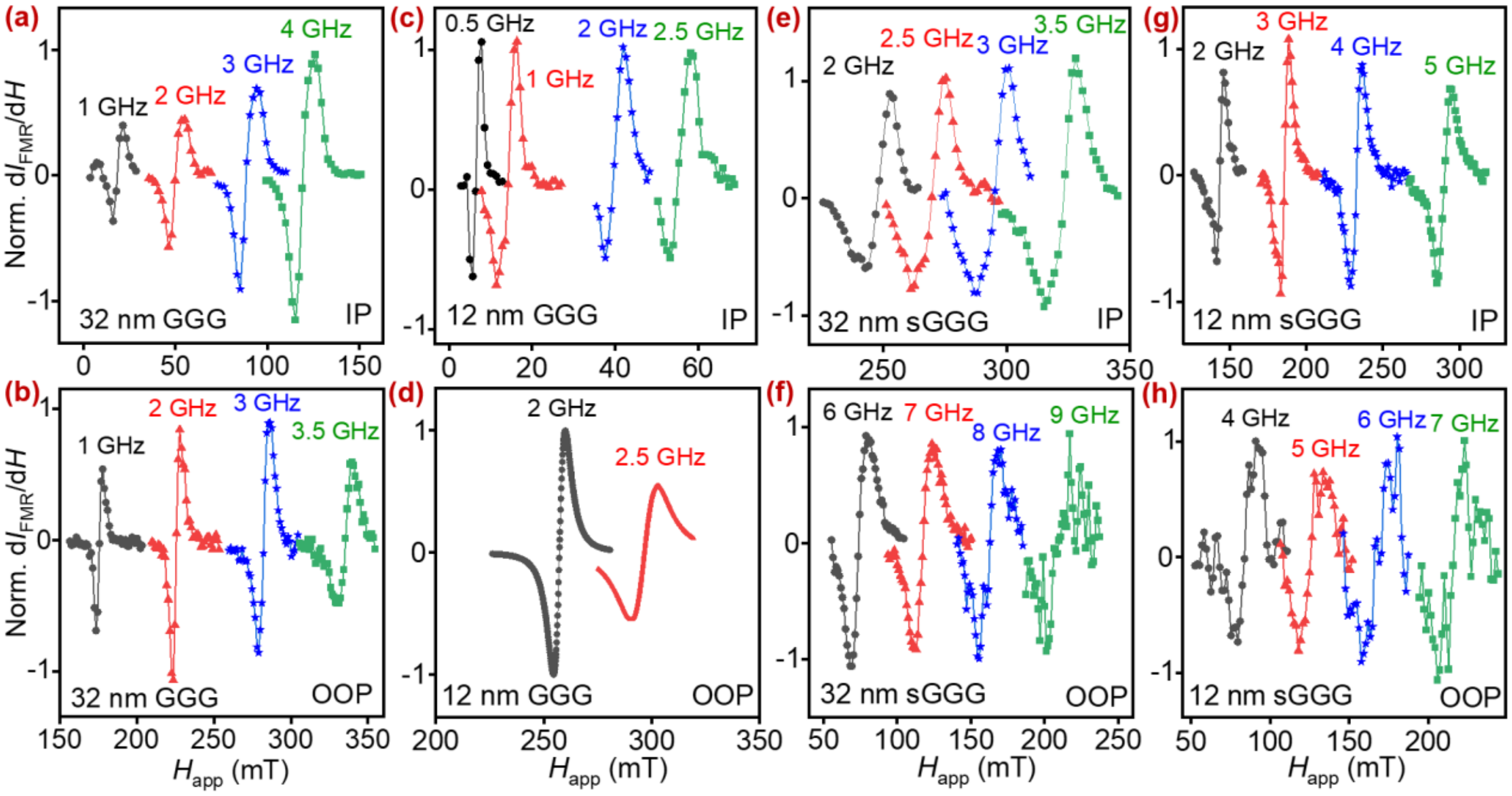

**Figure S4.** Normalized derivative FMR spectra of IP (a) and OOP (b) TmIG (32 nm)/GGG. Normalized derivative FMR spectra of IP (c) and OOP (d) TmIG (12 nm)/GGG. Normalized derivative FMR spectra of IP (e) and OOP (f) TmIG (32 nm)/sGGG. Normalized derivative FMR spectra of IP (g) and OOP (h) TmIG( 12 nm)/sGGG.

We performed PMOKE measurements on TmIG (7 nm)/GGG film (Figure S5a) and found a low saturation magnetic field (~30 mT). In addition, IP FMR was performed by measuring the $S_{22}$ at room temperature at different MW excitation frequencies, to extract the magnetic anisotropy and damping constant $\alpha$ of the TmIG (7 nm) /GGG. Figure S5b shows the FMR derivative spectra *vs* $H_{app}$ for various thicknesses, demonstrating distinct behaviors at different MW frequencies of 1 GHz, 2 GHz, 3 GHz, and 4 GHz. Figure S5c shows the measured (open circles) IP FMR resonance vs $H_{app}$ curves for TmIG (7 nm)/GGG film, fitted (solid lines) using the same procedures in references [33,39] leading to $\gamma$ of 22.4 GHz/T, $M_S$ of 66 kA/m (82.9 mT), in-plane magnetic anisotropy of 126.8 mT, and $\alpha$ of 0.0122. Further details on SW electrical transmission spectroscopy measurements are discussed below.

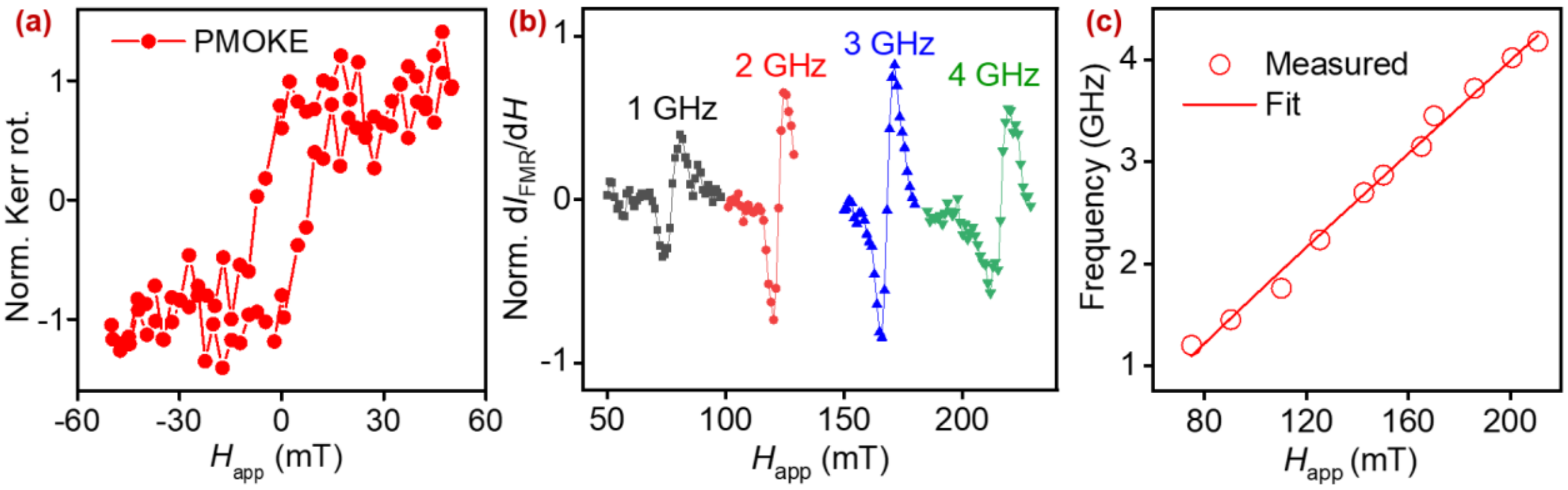

**Figure S5.** (a) Normalized Kerr rotation for PMOKE (filled circles) on TmIG (7 nm)/GGG film. (b) Normalized derivative FMR spectra of IP TmIG (7 nm)/GGG. (c) FMR IP measured (open circles) curve of TmIG (7 nm)/GGG fitted (solid line) to extract the magnetic properties of the film.

## S.3. Spin wave propagation properties for TmIG films

Spin-wave (SW) electrical transmission spectroscopy was performed in magnetostatic surface spin-wave (MSSW) geometry, *i.e.*, $H_{app}$ applied in-plane perpendicular to the SW proposition on TmIG (34 nm)/GGG films. Figure S6a shows IP real $S_{21}$ SW transmission intensity map (MW frequency *vs* $H_{app}$) at a MW power 1 mW. Selected imaginary $S_{21}$ spectra are plotted in Figure S6b at $H_{app}$ values of 10.65 mT, 20.5 mT, 42.6 mT, and 60.5 mT, respectively. $\delta f$ is the frequency difference between two closest maxima/minima of the Im$S_{21}$ (highlighted in the inset of Figure S6b), corresponding to a SW phase of $\pi$. The SW intensity of the other SW peaks (modes) can be optimized by choosing the distance $S$ between the ground and source signal in the coplanar waveguides (CPW) to excite them efficiently. The SW group velocity $v_g$ of MSSWs is plotted as function of $H_{app}$ (Figure S6c) with values ranging from 2 to 4 km/s. Figure S6d depicts the SW decay length $l_d$ *vs* $H_{app}$. $l_d$ up to 30 µm was measured at low magnetic fields (< 15 mT), in good agreement with previous nitrogen-vacancy (NV) magnetometry measurements done on the same film.[33]

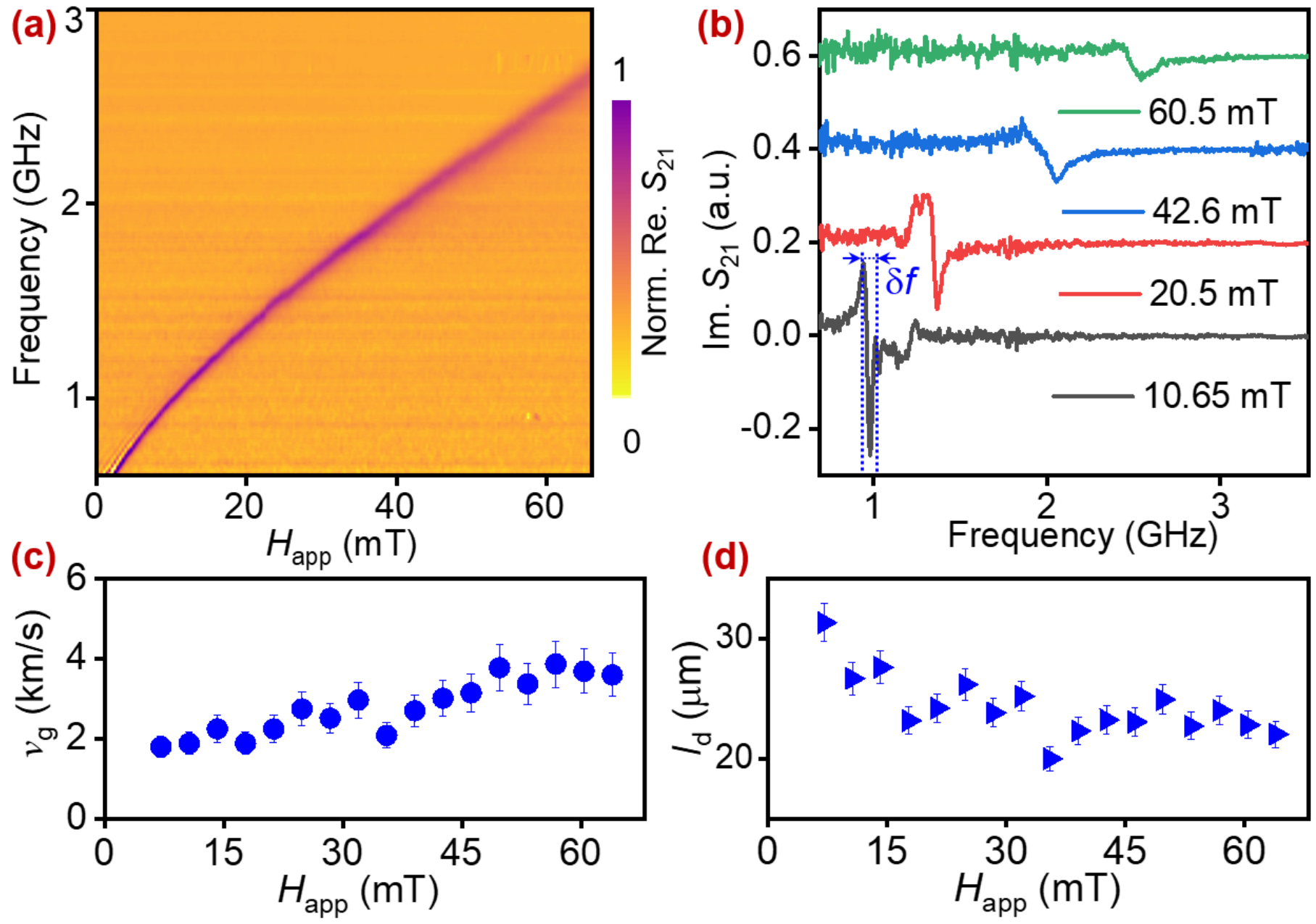

**Figure S6.** (a) SW $S_{21}$ intensity map (MW frequency *vs* $H_{app}$) for MSSW propagation on TmIG (34 nm)/GGG at a distance $S$ of 32 µm. (b) Imaginary $S_{21}$ spectra at 10.65 mT, 20.5 mT, 42.6mT, and 60.5 mT. SW group velocity (c) and decay length (d) of MSSWs as function of the applied magnetic field.

Figure S7 shows the SW $S_{21}$ intensity map (MW frequency *vs* $H_{app}$) of MSSWs at a MW power of 1 mW measured on TmIG (12 nm)/GGG film at a propagation distances $S$ of 32 µm, 40 µm, 60 µm, and 80 µm. As discussed in the main text, the SW amplitude decreases with a decay length of 30 µm, that agrees well with $l_d$ value at 20 mT deduced from Figure 4f.

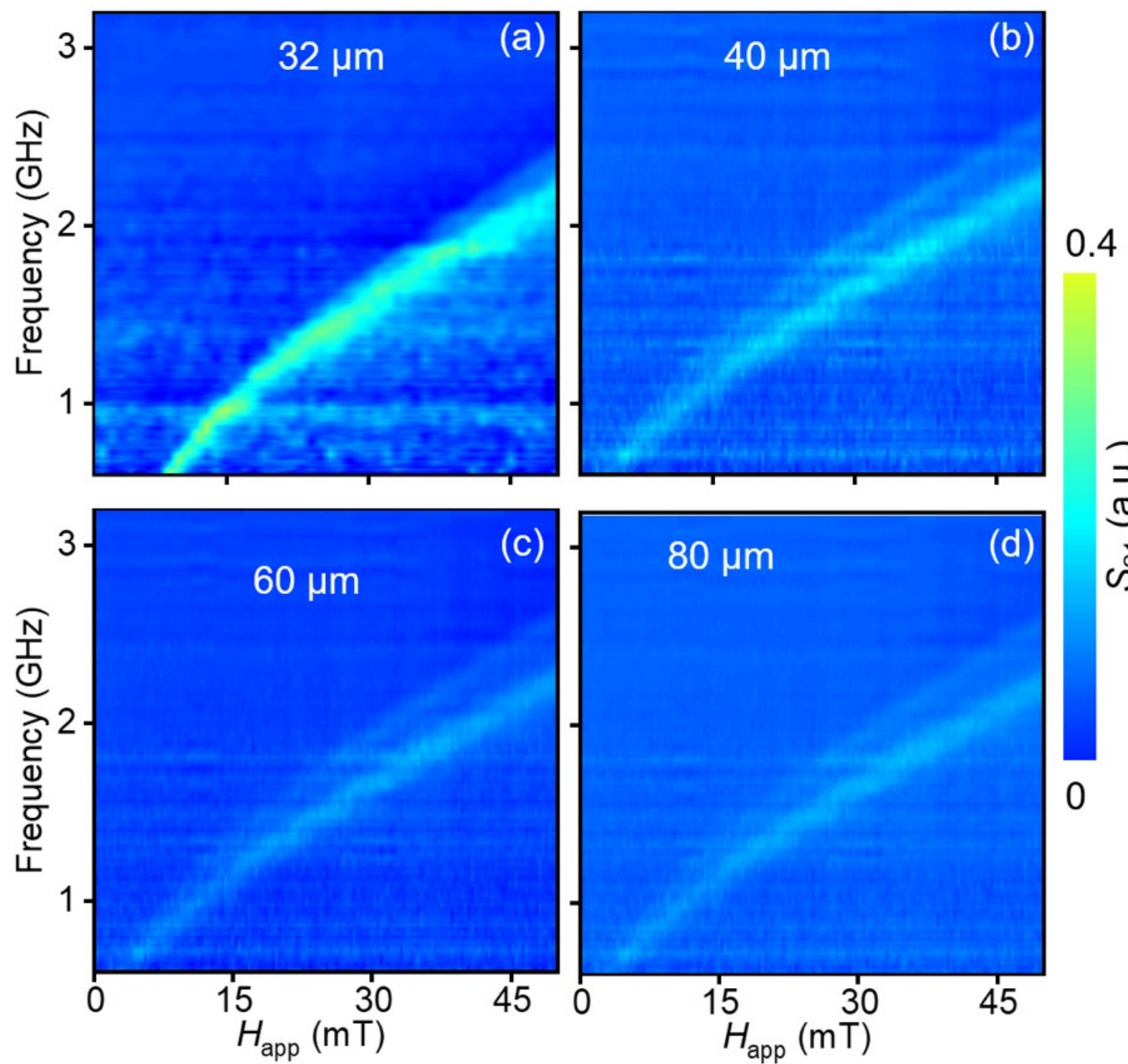

**Figure S7.** SW $S_{21}$ intensity map (MW frequency *vs* $H_{app}$) for the MSSW propagation on TmIG (12 nm)/GGG film at a distance $S$ of 32 µm (a), 40 µm (b), 60 µm (c), and 80 µm (d).

Figure S8a shows the IP real SW $S_{21}$ intensity map (MW frequency *vs* $H_{app}$) of TmIG (7 nm)/GGG film at a MW power of 10 mW. The real $S_{21}$ spectrum is plotted in Figure S8b at $H_{app}$ of 50 mT, showing propagating MSSW with a SW velocity $v_g$ in the range of 2 – 8 km/s, see Figure S7c. Due to the broad and weak signal of the FMR peaks, it was not possible to extract the SW decay length. Nanofabrciation of small (< 500 nm) MW CPW antennas may help in exciting such MSSWs effectively in ultrathin TmIG films.[14]

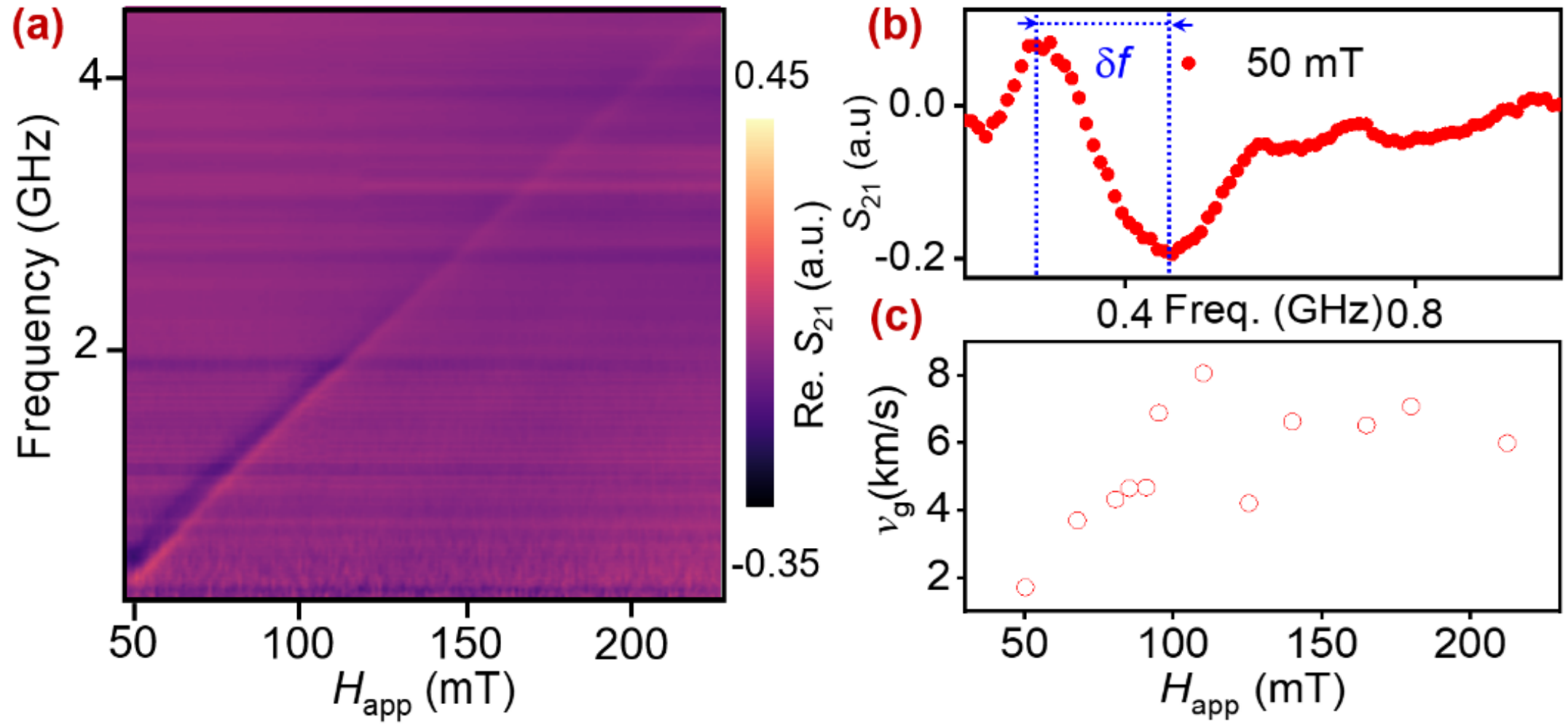

**Figure S8.** (a) SW $S_{21}$ intensity map (MW frequency *vs* $H_{app}$) of TmIG (7 nm)/GGG at a separation distance $S$ of 32 µm. (b) $S_{21}$ *vs* MW frequency spectrum at $H_{app}$ of 50.05 mT. (c) SW group velocity of MSSWs as function of $H_{app}$.

TmIG (32 nm)/GGG films also exhibit forward volume spin waves (FVSWs), suggesting OOP magnetic anisotropy.[32] Figure S9a shows SW real $S_{21}$ intensity map (MW frequency *vs* $H_{app}$) at a MW power of 1 mW measured in FVSW geometry. The real $S_{21}$ spectra are plotted in Figure S9b at $H_{app}$ of 167.56 mT and 211.58 mT, showing propagating FVSWs with group velocity $v_g$ of 2 – 4 km/s (see Figure S9c) and a decay length $l_d$ of 10 – 35 µm (see Figure S9d). The SW velocity of TmIG (32 nm)/GGG film is lower than the values measured (up to 8 km/s) on TmIG (7 nm)/GGG film that may be attributed to the broad SW signal (higher $\Delta f$) obtained at a high MW power of 10 mW.

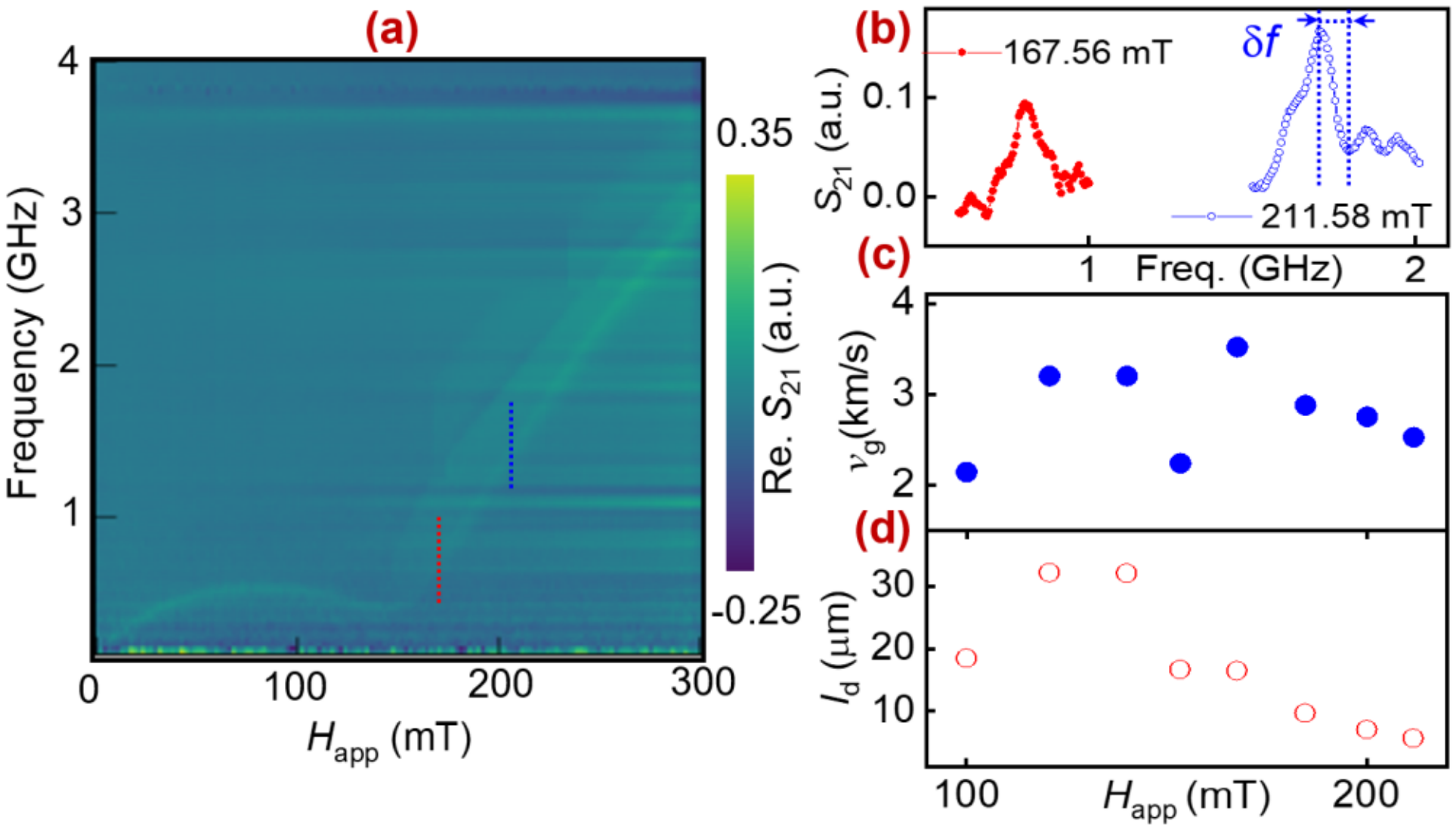

**Figure S9.** (a) SW $S_{21}$ intensity map (MW frequency *vs* $H_{app}$) for FVSW propagation on TmIG (32 nm)/GGG film at a separation CPW distance $S$ of 32 µm. (b) Real $S_{21}$ SW spectra at $H_{app}$ of 167.56 mT and 211.58 mT. FWSW group velocity (c) and decay length (d) as function of the applied magnetic field.

Finally, the nonreciprocity behavior of MSSWs in TmIG (32 nm)/GGG is also measured, as discussed in main text for the case of TmIG (12 nm)/GGG films. Figure S10 shows the nonreciprocity propagation of MSSW at two values of $H_{app}$ of 5.65 mT (Figure S10a) and 42.6 mT (Figure S10b). A strong single mode along with other propagating spin waves modes were observed for $S_{21}$ measurements while $S_{12}$ measurement showed no signal.

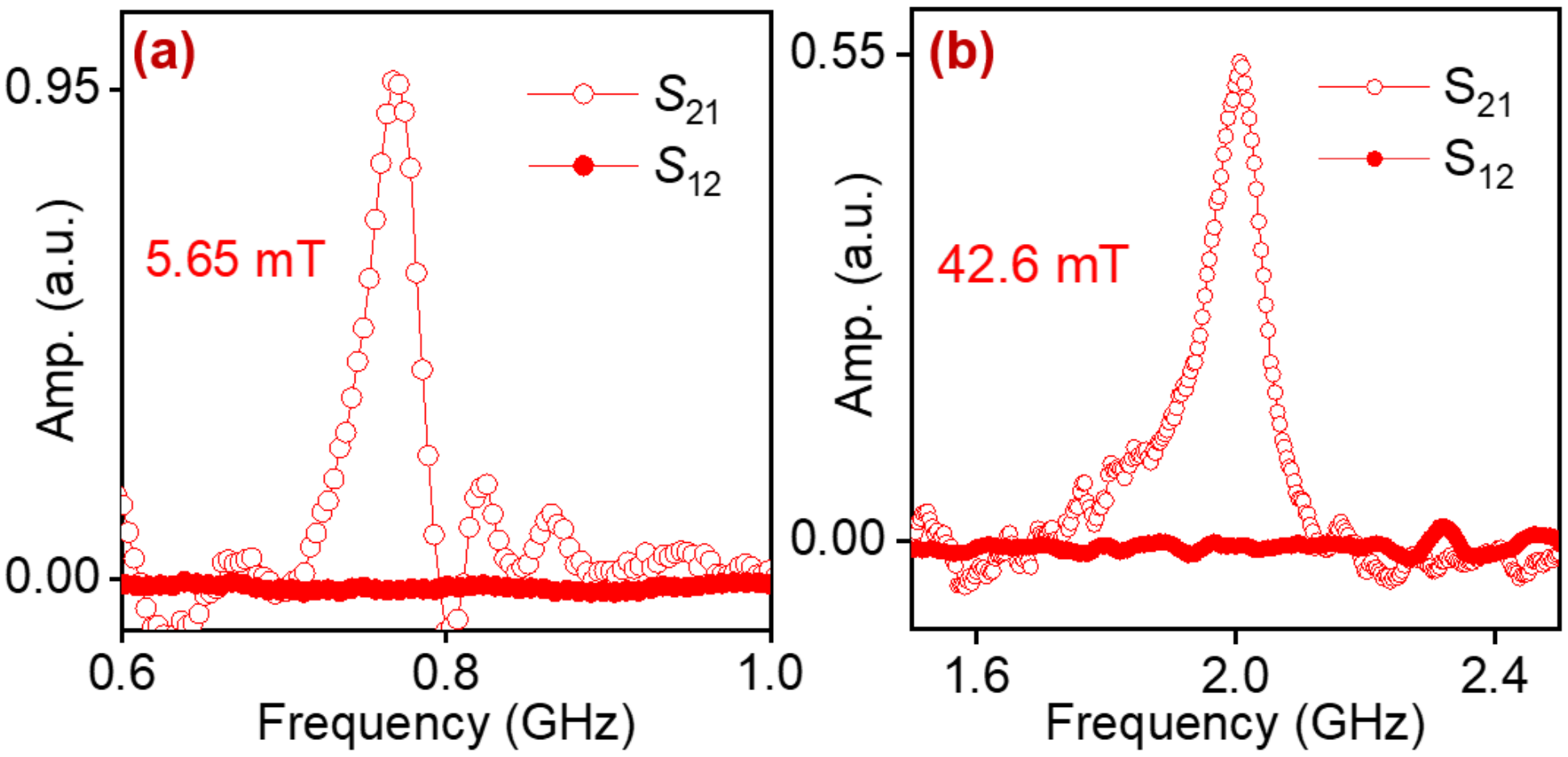

**Figure S10.** Nonreciprocity of MSSWs on TmIG (32 nm)/GGG at 5.65 mT (a) and 42.6 mT (b).

## References

[1] T. Sebastian, K. Schultheiss, B. Obry, B. Hillebrands, H. Schultheiss, *Front. Phys.* **2015**, *3*.

[2] H. Schultheiss, X. Janssens, M. van Kampen, F. Ciubotaru, S. J. Hermsdoerfer, B. Obry, A. Laraoui, A. A. Serga, L. Lagae, A. N. Slavin, B. Leven, B. Hillebrands, *Phys. Rev. Lett.* **2009**, *103*, 157202.

[3] L. Berger, *Phys. Rev. B* **1996**, *54*, 9353.

[4] M. Collet, X. de Milly, O. d'Allivy Kelly, V. V. Naletov, R. Bernard, P. Bortolotti, J. Ben Youssef, V. E. Demidov, S. O. Demokritov, J. L. Prieto, M. Muñoz, V. Cros, A. Anane, G. de Loubens, O. Klein, *Nat. Commun.* **2016**, *7*, 10377.

[5] P. Andrich, C. F. de las Casas, X. Liu, H. L. Bretscher, J. R. Berman, F. J. Heremans, P. F. Nealey, D. D. Awschalom, *Npj Quantum Inf.* **2017**, *3*, 1.

[6] I. Bertelli, J. J. Carmiggelt, T. Yu, B. G. Simon, C. C. Pothoven, G. E. W. Bauer, Y. M. Blanter, J. Aarts, T. Van Der Sar, *Sci. Adv.* **2020**, *6*, eabd3556.

[7] T. An, V. I. Vasyuchka, K. Uchida, A. V. Chumak, K. Yamaguchi, K. Harii, J. Ohe, M. B. Jungfleisch, Y. Kajiwara, H. Adachi, B. Hillebrands, S. Maekawa, E. Saitoh, *Nat. Mater.* **2013**, *12*, 549.

[8] A. Laraoui, V. Halté, M. Vomir, J. Vénuat, M. Albrecht, E. Beaurepaire, J.-Y. Bigot, *Eur. Phys. J. D* **2007**, *43*, 251.

[9] M. Vomir, L. H. F. Andrade, L. Guidoni, E. Beaurepaire, J.-Y. Bigot, *Phys. Rev. Lett.* **2005**, *94*, 237601.

[10] J. R. Hortensius, D. Afanasiev, M. Matthiesen, R. Leenders, R. Citro, A. V. Kimel, R. V. Mikhaylovskiy, B. A. Ivanov, A. D. Caviglia, *Nat. Phys.* **2021**, *17*, 1001.

[11] D. Kikuchi, D. Prananto, K. Hayashi, A. Laraoui, N. Mizuochi, M. Hatano, E. Saitoh, Y. Kim, C. A. Meriles, T. An, *Appl. Phys. Express* **2017**, *10*, 103004.

[12] K. An, V. S. Bhat, M. Mruczkiewicz, C. Dubs, D. Grundler, *Phys. Rev. Appl.* **2019**, *11*, 034065.

[13] A. V. Chumak, V. I. Vasyuchka, A. A. Serga, B. Hillebrands, *Nat. Phys.* **2015**, *11*, 453.

[14] J. Solano, O. Gladii, P. Kuntz, Y. Henry, D. Halley, M. Bailleul, *Phys. Rev. Mater.* **2022**, *6*, 124409.

[15] C. Liu, J. Chen, T. Liu, F. Heimbach, H. Yu, Y. Xiao, J. Hu, M. Liu, H. Chang, T. Stueckler, S. Tu, Y. Zhang, Y. Zhang, P. Gao, Z. Liao, D. Yu, K. Xia, N. Lei, W. Zhao, M. Wu, *Nat. Commun.* **2018**, *9*, 738.

[16] Y. Li, T.-H. Lo, J. Lim, J. E. Pearson, R. Divan, W. Zhang, U. Welp, W.-K. Kwok, A. Hoffmann, V. Novosad, *Appl. Phys. Lett.* **2023**, *123*, 022406.

[17] V. E. Demidov, S. O. Demokritov, K. Rott, P. Krzysteczko, G. Reiss, *J. Phys. Appl. Phys.* **2008**, *41*, 164012.

[18] A. B. Ustinov, B. A. Kalinikos, E. Lähderanta, *J. Appl. Phys.* **2013**, *113*, 113904.

[19] L. Zheng, L. Jin, T. Wen, Y. Liao, X. Tang, H. Zhang, Z. Zhong, *J. Phys. Appl. Phys.* **2022**, *55*, 263002.

[20] F. Ciubotaru, T. Devolder, M. Manfrini, C. Adelmann, I. P. Radu, *Appl. Phys. Lett.* **2016**, *109*, 012403.

[21] B. W. Zingsem, M. Winklhofer, R. Meckenstock, M. Farle, *Phys. Rev. B* **2017**, *96*, 224407.

[22] G. Varvaro, S. Laureti, D. Fiorani, *J. Magn. Magn. Mater.* **2014**, *368*, 415.

[23] L. Alahmed, P. Li, in *Magn. Mater. Magn. Levitation* (Eds.: D. Ranjan Sahu, V. N. Stavrou), IntechOpen, **2021**.

[24] O. Ciubotariu, A. Semisalova, K. Lenz, M. Albrecht, *Sci. Rep.* **2019**, *9*, 17474.

[25] A. Quindeau, C. O. Avci, W. Liu, C. Sun, M. Mann, A. S. Tang, M. C. Onbasli, D. Bono, P. M. Voyles, Y. Xu, J. Robinson, G. S. D. Beach, C. A. Ross, *Adv. Electron. Mater.* **2017**, *3*, 1600376.

[26] C. N. Wu, C. C. Tseng, Y. T. Fanchiang, C. K. Cheng, K. Y. Lin, S. L. Yeh, S. R. Yang, C. T. Wu, T. Liu, M. Wu, M. Hong, J. Kwo, *Sci. Rep.* **2018**, *8*, 11087.

[27] L. Caretta, E. Rosenberg, F. Büttner, T. Fakhrul, P. Gargiani, M. Valvidares, Z. Chen, P. Reddy, D. A. Muller, C. A. Ross, G. S. D. Beach, *Nat. Commun.* **2020**, *11*, 1090.

[28] C. O. Avci, E. Rosenberg, L. Caretta, F. Büttner, M. Mann, C. Marcus, D. Bono, C. A. Ross, G. S. D. Beach, *Nat. Nanotechnol.* **2019**, *14*, 561.
[29] A. S. Ahmed, A. J. Lee, N. Bagués, B. A. McCullian, A. M. A. Thabt, A. Perrine, P.-K. Wu, J. R. Rowland, M. Randeria, P. C. Hammel, D. W. McComb, F. Yang, *Nano Lett.* **2019**, *19*, 5683.
[30] Q. Shao, C. Tang, G. Yu, A. Navabi, H. Wu, C. He, J. Li, P. Upadhyaya, P. Zhang, S. A. Razavi, Q. L. He, Y. Liu, P. Yang, S. K. Kim, C. Zheng, Y. Liu, L. Pan, R. K. Lake, X. Han, Y. Tserkovnyak, J. Shi, K. L. Wang, *Nat. Commun.* **2018**, *9*, 3612.
[31] L. Sheng, Y. Liu, J. Chen, H. Wang, J. Zhang, M. Chen, J. Ma, C. Liu, S. Tu, C.-W. Nan, H. Yu, *Appl. Phys. Lett.* **2020**, *117*, 232407.
[32] J. Xu, D. Zhang, Y. Zhang, Z. Zhong, H. Zhang, X. Xu, X. Luo, Q. Yang, B. Liu, L. Jin, *AIP Adv.* **2022**, *12*, 065026.
[33] R. Timalsina, H. Wang, B. Giri, A. Erickson, X. Xu, A. Laraoui, *Adv. Electron. Mater.* **2024**, *10*, 2300648.
[34] J. Čermák, A. Abrahám, T. Fabián, P. Kaboš, P. Hyben, *J. Magn. Magn. Mater.* **1990**, *83*, 427.
[35] N. S. Rajput, K. Sloyan, D. H. Anjum, M. Chiesa, A. A. Ghaferi, *Ultramicroscopy* **2022**, *235*, 113496.
[36] G. Greczynski, L. Hultman, *J. Appl. Phys.* **2022**, *132*, 011101.
[37] S. Lamichhane, R. Timalsina, C. Schultz, I. Fescenko, K. Ambal, S.-H. Liou, R. Y. Lai, A. Laraoui, *Nano Lett.* **2024**, acs.nanolett.3c03843.
[38] R. Sharma, P. K. Ojha, S. Choudhary, S. K. Mishra, *Mater. Lett.* **2023**, *352*, 135154.
[39] A. J. Lee, A. S. Ahmed, B. A. McCullian, S. Guo, M. Zhu, S. Yu, P. M. Woodward, J. Hwang, P. C. Hammel, F. Yang, *Phys. Rev. Lett.* **2020**, *124*, 257202.
[40] C. Tang, P. Sellappan, Y. Liu, Y. Xu, J. E. Garay, J. Shi, *Phys. Rev. B* **2016**, *94*, 140403.
[41] S. Crossley, A. Quindeau, A. G. Swartz, E. R. Rosenberg, L. Beran, C. O. Avci, Y. Hikita, C. A. Ross, H. Y. Hwang, *Appl. Phys. Lett.* **2019**, *115*, 172402.
[42] A. Erickson, S. Q. Abbas Shah, A. Mahmood, I. Fescenko, R. Timalsina, C. Binek, A. Laraoui, *RSC Adv.* **2023**, *13*, 178.
[43] B. G. Simon, S. Kurdi, J. J. Carmiggelt, M. Borst, A. J. Katan, T. van der Sar, *Nano Lett.* **2022**, *22*, 9198.
[44] S. Klingler, P. Pirro, T. Brächer, B. Leven, B. Hillebrands, A. V. Chumak, *Appl. Phys. Lett.* **2015**, *106*, 212406.
[45] J. Chen, H. Yu, G. Gubbiotti, *J. Phys. Appl. Phys.* **2022**, *55*, 123001.
[46] M. Jamali, J. H. Kwon, S.-M. Seo, K.-J. Lee, H. Yang, *Sci. Rep.* **2013**, *3*, 3160.
[47] N. Tang, W. L. N. C. Liyanage, S. A. Montoya, S. Patel, L. J. Quigley, A. J. Grutter, M. R. Fitzsimmons, S. Sinha, J. A. Borchers, E. E. Fullerton, L. DeBeer-Schmitt, D. A. Gilbert, *Adv. Mater.* **2023**, *35*, 2300416.
[48] P. A. McClarty, *Annu. Rev. Condens. Matter Phys.* **2022**, *13*, 171.
[49] A. Laraoui, K. Ambal, *Appl. Phys. Lett.* **2022**, *121*, 060502.
[50] T. X. Zhou, J. J. Carmiggelt, L. M. Gächter, I. Esterlis, D. Sels, R. J. Stöhr, C. Du, D. Fernandez, J. F. Rodriguez-Nieva, F. Büttner, E. Demler, A. Yacoby, *Proc. Natl. Acad. Sci.* **2021**, *118*, e2019473118.
[51] D. D. Awschalom, C. R. Du, R. He, F. J. Heremans, A. Hoffmann, J. Hou, H. Kurebayashi, Y. Li, L. Liu, V. Novosad, J. Sklenar, S. E. Sullivan, D. Sun, H. Tang, V. Tyberkevych, C. Trevillian, A. W. Tsen, L. R. Weiss, W. Zhang, X. Zhang, L. Zhao, Ch. W. Zollitsch, *IEEE Trans. Quantum Eng.* **2021**, *2*, 1.
[52] D. Xu, X.-K. Gu, H.-K. Li, Y.-C. Weng, Y.-P. Wang, J. Li, H. Wang, S.-Y. Zhu, J. Q. You, *Phys. Rev. Lett.* **2023**, *130*, 193603.
[53] M. Fukami, D. R. Candido, D. D. Awschalom, M. E. Flatté, *PRX Quantum* **2021**, *2*, 040314.
[54] *Acta Virol.* **1975**, *19*, 510.